\documentclass[useAMS]{gGAF2e}

\newcommand{\bA}{{\bf A}}
\newcommand{\bbb}{{\bf b}}
\newcommand{\bB}{{\bf B}}
\newcommand{\bD}{{\bf D}}
\newcommand{\bscE}{\bm {\cal{E}}}
\newcommand{\bF}{{\bf F}}
\newcommand{\bg}{{\bf g}}
\newcommand{\bG}{{\bf G}}
\newcommand{\bJ}{{\bf J}}
\newcommand{\bk}{{\bf k}}
\newcommand{\br}{{\bf r}}
\newcommand{\bR}{{\bf R}}
\newcommand{\bu}{{\bf u}}
\newcommand{\bU}{{\bf U}}
\newcommand{\bV}{{\bf V}}
\newcommand{\bW}{{\bf W}}
\newcommand{\bz}{{\bf z}}
\newcommand{\balpha}{{\bm \alpha}}
\newcommand{\bbeta}{{\bm \beta}}
\newcommand{\bgamma}{{\bm \gamma}}
\newcommand{\bdelta}{{\bm \delta}}
\newcommand{\bkappa}{{\bm \kappa}}
\newcommand{\bvarphi}{{\bm \varphi}}
\newcommand{\bOmega}{{\bm \Omega}}
\newcommand{\bmB}{\overline{{\bf B}}}
\newcommand{\bmU}{\overline{{\bf U}}}
\newcommand{\mB}{\overline{B}}
\newcommand{\bnab}{{\mbox{\boldmath $\nabla$}}}
\def\bzo {{\bf 0}}
\def\dd {\mbox{d}}

\def\iu {\mbox{i}}
\def\p {\partial}
\def\x {\times}
\def\ol {\overline}

\begin{document}

\doi{10.1080/03091920xxxxxxxxx} \issn{1029-0419} \issnp{0309-1929}
 \jvol{00} \jnum{00} \jyear{2006} \jmonth{February}

\markboth{Karl-Heinz R\"{a}dler \& Rodion Stepanov}{On the effects
of turbulence on a screw dynamo}

\title{On the effects of turbulence on a screw dynamo}
\author{KARL-HEINZ R\"{A}DLER * $\dagger$ and RODION STEPANOV $\ddagger$ \\
$\dagger$ Astrophysikalisches Institut Potsdam, An der Sternwarte
16, D-14482 Potsdam, Germany \\
$\ddagger$ Institute of Continuous Media Mechanics, Korolyov Street
1, Perm, 614013, Russia \thanks{ \vspace{6pt}
\newline{\tiny{ {\em } * Corresponding author. E-mail khraedler@arcor.de }}}}
\received{Received 12 December 2005; in final form 7 March 2006}

\maketitle

\begin{abstract}
In an experiment in the Institute of Continuous Media Mechanics in
Perm (Russia) an non--stationary screw dynamo is intended to be
realized with a helical flow of liquid sodium in a torus. The flow
is necessarily turbulent, that is, may be considered as a mean
flow and a superimposed turbulence. In this paper the induction
processes of the turbulence are investigated within the framework
of mean--field electrodynamics. They imply of course a part which
leads to an enhanced dissipation of the mean magnetic field. As a
consequence of the helical mean flow there are also helical
structures in the turbulence. They lead to some kind of
$\alpha$--effect, which might basically support the screw dynamo.
The peculiarity of this $\alpha$--effect explains measurements
made at a smaller version of the device envisaged for the dynamo
experiment. The helical structures of the turbulence lead also to
other effects, which in combination with a rotational shear are
potentially capable of dynamo action. A part of them can basically
support the screw dynamo. Under the conditions of the experiment
all induction effects of the turbulence prove to be rather weak in
comparison to that of the main flow. Numerical solutions of the
mean--field induction equation show that all the induction effects
of the turbulence together let the screw dynamo threshold
slightly, at most by one per cent, rise. The numerical results
give also some insights into the action of the individual
induction effects of the turbulence. 
\end{abstract}

\section{Introduction}

A screw--like, that is, helical motion of an electrically conducting
fluid is capable of dynamo action. This has been first theoretically
shown by \citet{ponomarenko73}. Later, in 1999, it was
experimentally demonstrated in the liquid sodium facility of the
Institute for Physics in Riga (Latvia)
\citep{gailitisetal00,gailitisetal01,gailitisetal01b,gailitisetal02,gailitisetal02b,gailitisetal03,gailitisetal04}.
It is also the background of another liquid sodium experiment which
is under preparation in the Institute for Continuous Media Mechanics
in Perm (Russia) (Denisov {\it et al.} 1999, Frick {\it et al.}
2001, 2002) \citep{denisovetal99,fricketal01,fricketal02}.

In this experiment a torus-shaped vessel filled with liquid sodium is rotated
about its symmetry axis and then suddenly stopped.
After the stop a flow of sodium occurs inside the vessel and decays
in the course of time.
Due to diverters, that is, proper arrangements of blades inside the vessel,
the flow becomes helical.
In this way the possibility of a dynamo of Ponomarenko type arises for
a certain time interval.
The large radius of the torus is about 0.40 m
and the small one about 0.12 m.
Rotation rates up to 50 rps are envisaged.
Estimates show that magnetic Reynolds numbers sufficient for dynamo action
are indeed feasible.

The critical magnetic Reynolds number is definitely higher than 10.
For liquid sodium the hydrodynamic Reynolds number exceeds
the magnetic Reynolds number by a factor of about $10^5$.
Therefore in the case of a dynamo the fluid flow is necessarily turbulent,
that is, may be considered as consisting of a main flow and a superimposed turbulence.
The turbulence affects the dynamo in two ways.
Firstly it influences the profile of the main flow.
Secondly it leads to small--scale induction effects.
In particular the turbulence will enlarge the effective magnetic diffusivity
of the fluid, what raises the dynamo threshold.
Likewise other small--scale induction effects will occur.
As a consequence of helical features of the turbulence an $\alpha$--effect
or related effects are possible.
Basically they could also support the screw dynamo or even open the possibility of another type of dynamo.

In this paper we want to estimate these direct influences of
turbulence on the electromagnetic field on the basis of
mean--field electrodynamics. We will rely on a recent general
calculation of the mean electromotive force due to turbulence
\citep{raeste05}, which was partially motivated by this
experiment. In section \ref{concept} we describe the general
concept of our investigation. In section \ref{structure} we
discuss possible structures of the mean electromotive force which
can be concluded from symmetry arguments, and in section
\ref{results} we explain results of more detailed calculations. On
this basis we deliver in section \ref{estimates} estimates for the
influence of the turbulence on the dynamo. Finally in section
\ref{summary} the main results are summarized.

\section{The mean--field concept}
\label{concept}

Let us first recall some ideas of the mean--field concept which are
important for the following \citep[see,
e.g.,][]{krauseetal71,moffatt78,raedler00}. We start from the
induction equation that governs the behavior of the magnetic field
$\bB$ in an electrically conducting fluid,
\begin{equation}
\partial_t \bB - \bnab \x (\bU \x \bB) - \eta \bnab^2 \bB = \bzo \, , \quad
   \bnab \cdot \bB = 0  \, .
\label{eq03}
\end{equation}
Here $\bU$ is the velocity of the fluid and $\eta$ its magnetic diffusivity,
which is assumed to be constant.

The velocity $\bU$ and, as a consequence, also the magnetic field
$\bB$ are assumed to consist of large--scale parts and small--scale
turbulent fluctuations. We define the mean magnetic and velocity
fields $\bmB$ and $\bmU$ as averages over space or time scales
larger than those of the turbulence and put $\bB = \bmB + \bbb$
analogously to $\bU = \bmU + \bu$. We assume that the Reynolds
averaging rules apply. Taking the average of equation (\ref{eq03})
we obtain the mean-field induction equation
\begin{equation}
\partial_t \bmB - \bnab \x (\bmU \x \bmB + \bscE)
   - \eta \bnab^2 \bmB = \bzo \, , \quad
   \bnab \cdot \bmB = 0  \, ,
\label{eq05}
\end{equation}
where $\bscE$ is an electromotive force due to the fluctuations of magnetic field
and velocity,
\begin{equation}
\bscE = \ol{\bu \x \bbb} \, . \label{eq07}
\end{equation}

The equation for $\bbb$ resulting from (\ref{eq03}) and (\ref{eq05})
allows us to conclude that $\bscE$ is determined by $\bmU$, $\bu$
and $\bmB$. More precisely, $\bscE$ at a given point in space and
time depends not only on $\bmU$, $\bu$ and $\bmB$ at this point but
also on their behavior in a certain neighborhood of this point. We
adopt the frequently used assumption that $\bmB$ varies only weakly
in space and time so that $\bscE$ in a given point depends on $\bmB$
only via its components and their first spatial derivatives in this
point. Then $\bscE$ can be represented in the form
\begin{equation}
{\cal{E}}_i = a_{ij} \, \overline{B}_j
 + b_{ijk} \, \partial \overline{B}_j / \partial x_k \, ,
\label{eq09}
\end{equation}
with tensors $a_{ij}$ and $b_{ijk}$ being averaged quantities determined
by $\bmU$ and $\bu$.
Here and in the following we refer to a Cartesian co--ordinate system
$(x_1, x_2, x_3)$ and adopt the summation convention.
Relation (\ref{eq09}) is equivalent to
\begin{eqnarray}
\bscE  &=& - \balpha \circ \bmB - \bgamma \x \bmB
\nonumber\\
&& - \bbeta \circ (\bnab \x \bmB) - \bdelta \x (\bnab \x \bmB)
   - \bkappa \circ (\bnab \bmB)^{(s)}
\label{eq11}
\end{eqnarray}
(see R\"adler 1980, 2000), where $\balpha$ and $\bbeta$ are
symmetric tensors of the second rank, $\bgamma$ and $\bdelta$ are
vectors, and $\bkappa$ is a tensor of the third rank, all
depending on $\bmU$ and $\bu$ only. $(\bnab \bmB)^{(s)}$ is the
symmetric part of the gradient tensor of $\bmB$, i.e. $(\bnab
\bmB)^{(s)}_{jk} = \frac{1}{2} (\partial \ol{B}_j /
\partial x_k + \partial \ol{B}_k / \partial x_j)$.
Notations like $\balpha \circ \bmB$ or $\bkappa \circ (\bnab
\bmB)^{(s)}$ are used in the sense of $(\balpha \circ \bmB)_i =
\alpha_{ij} {\ol{B}}_j$ or $(\bkappa \circ (\bnab \bmB)^{(s)})_i =
\kappa_{ijk} ((\bnab \bmB)^{(s)})_{jk}$. The term with $\balpha$
in (\ref{eq11}) describes the $\alpha$--effect, which is in
general anisotropic, that with $\bgamma$ a transport of mean
magnetic flux by the turbulence. The terms with $\bbeta$ and
$\bdelta$ can be interpreted by introducing a modified magnetic
diffusivity, again in general anisotropic. They are usually
accompanied by the $\bkappa$ term, which allows no simple
interpretation.

\section{The structure of the mean electromotive force $\bscE$
in the presence of a screw motion}
\label{structure}

\subsection{Mean motion}

We consider the Ponomarenko dynamo here in its original cylindrical
geometry. As observed in other investigations this provides a
reasonable approximation for the situation in a torus even for
aspect ratios as large as in the planned device \citep{stepanov00}.
We assume that it is primarily a helical mean flow described by the
mean velocity $\bmU$ which generates or maintains the mean magnetic
field $\bmB$. Referring to a proper cylindrical co--ordinate system
$(r, \varphi, z)$ we put
\begin{equation}
\bmU = ( 0, {\ol{U}}_\varphi (r), {\ol{U}}_z (r) ) \, , \quad
    {\ol{U}}_\varphi = \Omega (r) r \, .
\label{eq01}
\end{equation}
As mentioned above we will study the influence of the turbulence, that is, of the velocity field $\bu$,
on the mean magnetic field $\bmB$ in the framework of mean--field electrodynamics.

For a more detailed discussion of the mean electromotive force $\bscE$ caused by the turbulence
and for its calculation we change temporarily from the frame of reference in which (\ref{eq01}) applies
to another one rotating about the axis $r = 0$ and moving along it.
Let us consider $\bscE$ in a given point at the surface $r = r_0$.
The moving frame is then fixed such that the mean velocity of the fluid
at the surface $r = r_0$ is equal to zero.
The result for $\bscE$ found in this rotating system,
considered as a vector with the usual transformation properties,
applies in the original frame of reference as well.
In the moving frame the mean velocity $\bmU$ has,
again in the cylindrical co--ordinate system introduced above,
the form
\begin{equation}
\bmU = ( 0, (\Omega (r) - \Omega (r_0)) r, \ol{U}_z (r) - \ol{U}_z (r_0) ) \, .
\label{eq15}
\end{equation}
In contrast to the original frame, in the rotating one a Coriolis
force occurs, which is defined by an angular velocity $\bOmega$,
given by
\begin{equation}
\bOmega = \Omega (r_0) \, \hat{\bz} \, ,
\label{eq17}
\end{equation}
where $\hat{\bz}$ means the unit vector in the $z$ direction
of the cylindrical co--ordinate system.

\subsection{Homogeneous turbulence}

Consider now $\bscE$ in the moving frame at a point with $\bmU =
\bzo$. Then $\bscE$ and so the  quantities $a_{ij}$ and $b_{ijk}$,
or $\balpha$, $\bgamma$, $\bbeta$, $\bdelta$ and $\bkappa$, depend
on $\bmU$ and $\bu$ in a certain neighborhood of this point. The
dependence on $\bu$ implies again a dependence on $\bmU$ and of
course also a dependence on $\bOmega$. We restrict our attention
now to the case in which $\bmU$ varies only weakly in the relevant
neighborhood and assume that its behavior can be there
sufficiently precisely described by the gradient tensor $\bnab
\bmU$, with respect to Cartesian co--ordinates defined by $(\bnab
\bmU)_{ij} = \partial {\ol{U}}_i / \partial x_j$. In addition we
assume until further notice that the turbulent fluctuations $\bu$
deviate from a homogeneous isotropic mirror--symmetric turbulence
only as a consequence of the gradient $\bnab \bmU$ of $\bmU$ and
the Coriolis force defined by $\bOmega$. We may split $\bnab \bmU$
into its symmetric and antisymmetric parts. The first one is the
rate of strain tensor $\bD$, that is, $D_{ij} = \frac{1}{2}
(\partial {\ol{U}}_i / \partial x_j + \partial {\ol{U}}_j /
\partial x_i)$. It describes the deforming motion near the point
considered. Due to the incompressibility of the fluid we have
$\bnab \cdot \bmU = 0$ and therefore $D_{ii} = 0$. The second
part, $\bA$, given by $A_{ij} = \frac{1}{2} (\partial {\ol{U}}_i /
\partial x_j - \partial {\ol{U}}_j / \partial x_i)$, corresponds
to a rigid body rotation of the fluid near this point. We may
represent it according to $A_{ij} = - \frac{1}{2}\epsilon_{ijl}
W_l$ by a vector $\bW = \bnab \x \bmU$. Until further notice we do
not use specifications of $\bD$ and $\bW$ according to
(\ref{eq15}). They will be introduced only later.

We now utilize the symmetry properties of the basic equations
governing $\bB$ and $\bU$, or $\bmB$, $\bmU$, $\bbb$ and $\bu$, from
which $\bscE$ and so the quantities $a_{ij}$ and $b_{ijk}$, or
$\balpha$, $\bgamma$, $\bbeta$, $\bdelta$ and $\bkappa$, have to be
derived. They allow two important conclusions \citep[see also,
e.g.,][]{raeste05}. The first reads that these quantities cannot
contain any other construction elements than the isotropic tensors
$\delta_{lm}$ and $\epsilon_{lmn}$, the vectors $\bOmega$ and $\bW$
and the tensor $\bD$. The second uses the distinction between polar
and axial vectors as $\bU$ and $\bB$, respectively, and between true
and pseudo tensors. It reads that $a_{ij}$ and $b_{ijk}$, further
$\balpha$ and $\bkappa$ are pseudo tensors, $\bbeta$ is a true
tensor, $\bgamma$ a polar vector, and $\bdelta$ an axial vector. It
is then important that $\bOmega$ and $\bW$ are axial vectors and
$\bD$ is a true tensor.

Let us consider first $\alpha_{ij}$ and $\gamma_i$.
The mentioned construction elements $\delta_{lm}$, $\epsilon_{lmn}$, $\bOmega$, $\bW$ and $\bD$
allow us neither to built a pseudo tensor of the second rank nor a polar vector.
That is, we have
\begin{equation}
\alpha_{ij} = 0 \, , \quad \gamma_i = 0 \, .
\label{eq19}
\end{equation}
We stress that this applies independently of any assumptions concerning linearity
in $\bOmega$, $\bW$ and $\bD$ like those we will introduce below.

In contrast to this there are several non--zero contributions to $\beta_{ij}$,
$\delta_i$ and $\kappa_{ijk}$.
For the sake of simplicity we give here only those of them which are linear
in $\bOmega$, $\bW$ and $\bD$,
\begin{eqnarray}
\beta_{ij} &=& \beta^{(0)} \delta_{ij} + \beta^{(D)} D_{ij} \, , \quad
    \delta_i = \delta^{(\Omega)} \Omega_i + \delta^{(W)} W_i  \, , \quad
\nonumber\\
\kappa_{ijk} &=& \frac{1}{2} \kappa^{(\Omega)} (\Omega_j \delta_{ik} + \Omega_k \delta_{ij})
    + \frac{1}{2} \kappa^{(W)} (W_j \delta_{ik} + W_k \delta_{ij})
    + \kappa^{(D)} (\epsilon_{ijl} D_{kl} + \epsilon_{ikl} D_{jl}) \, .
\label{eq21}
\end{eqnarray}
The coefficients $\beta^{(0)}$, $\beta^{(D)}$,
$\delta^{(\Omega)}$, $\cdots$ $\kappa^{(D)}$ are determined by
$\bu$ but do not depend on $\bOmega$, $\bW$ and $\bD$. Because of
$\bnab \cdot \bmB = 0$, terms of $\kappa_{ijk}$ containing
$\delta_{jk}$ would not contribute to $\bscE$ and have therefore
been dropped.

As a consequence of (\ref{eq19}) and (\ref{eq21}) we have
\begin{eqnarray}
\bscE  &=& - \beta^{(0)} \bnab \x \bmB
    - \beta^{(D)} \bD \circ (\bnab \x \bmB)
\nonumber\\
&& - (\delta^{(\Omega)} \bOmega + \delta^{(W)} \bW) \x (\bnab \x \bmB)
\label{eq23}\\
&& - (\kappa^{(\Omega)} \bOmega + \kappa^{(W)} \bW) \circ (\bnab \bmB)^{(s)}
   - \kappa^{(D)} \, \hat{\bkappa} (\bD) \circ (\bnab \bmB)^{(s)} \, ,
\nonumber
\end{eqnarray}
where $\hat{\bkappa} (\bD)$ is a tensor of the third rank
defined by ${\hat{\kappa}}_{ijk} = \epsilon_{ijl} D_{lk} + \epsilon_{ikl} D_{lj}$.

The $\beta^{(0)}$ and $\beta^{(D)}$ terms in (\ref{eq23}) make
that the mean--field diffusivity deviates from the original, that
is, molecular magnetic diffusivity of the fluid. Due to the
$\beta^{(D)}$ term the mean--field diffusivity becomes
anisotropic. The $\delta^{(\Omega)}$ and $\delta^{(W)}$ terms,
too, can be discussed as contributions to the mean--field
diffusivity. They lead to skew--symmetric contributions to the
diffusivity tensor. In another context the effect described by the
$\delta^{(\Omega)}$ term has been called ``$\bOmega \x
\bJ$--effect". It has been shown that this effect in combination
with a differential rotation, here a dependence of $\Omega$ or
$r$, is able to establish a dynamo
\citep{raedler69,roberts72,moffattetal82,raedleretal03}. The
$\delta^{(W)}$ term describes an effect analogous to the $\bOmega
\x \bJ$--effect, which has been revealed only recently
\citep{rogachevskiietal03,raeste05}. It occurs even in the absence
of the Coriolis force as consequence of a shear in the mean
motion. The possibility of a dynamo due to this effect is still
under debate, see section~\ref{subsec53}. We refrain from
discussing the $\kappa^{(\Omega)}$, $\kappa^{(W)}$ and
$\kappa^{(D)}$ terms here.

\subsection{Inhomogeneous turbulence}

Let us now relax the assumption that the original turbulence is homogeneous and isotropic.
We admit an inhomogeneity and an anisotropy due to a gradient
of the turbulence intensity and introduce a vector $\bg$
in the direction of this gradient by putting $\bnab \ol{u^2} = \bg \ol{u^2}$.
Then we have to add $\bg$, which is a polar vector, to the above--mentioned construction elements
of $\alpha_{ij}$, $\gamma_i$, $\beta_{ij}$, $\delta_i$ and $\kappa_{ijk}$.
As a consequence $\balpha$ and $\bgamma$ can well be non-zero.
For the sake of simplicity we assume that the influence of $\bg$
on $\balpha$, $\bbeta$, $\bgamma$, $\bdelta$ and $\bkappa$ is so weak
that they are at most of first order in $\bg$.
We have then
\begin{eqnarray}
\alpha_{ij}  &=& \alpha_1^{(\Omega)} (\bg \cdot \bOmega) \delta_{ij}
    + \alpha_2^{(\Omega)} (g_i \Omega_j + g_j \Omega_i)
    + \alpha_1^{(W)} (\bg \cdot \bW) \delta_{ij}
    + \alpha_2^{(W)} (g_i W_j + g_j W_i)
\nonumber\\
&& \quad
    + \alpha^{(D)} (\epsilon_{ilm} D_{jl} + \epsilon_{jlm} D_{il}) \, g_m
\label{eq25}\\
\gamma_i &=& \gamma^{(0)} g_i
    + \gamma^{(\Omega)} \epsilon_{ilm} g_l \Omega_m
    + \gamma^{(W)} \epsilon_{ilm} g_l W_m
    + \gamma^{(D)} g_j D_{ij} \, ,
\nonumber
\end{eqnarray}
whereas (\ref{eq21}) remains unchanged.

The general form of $\bscE$ reads now
\begin{eqnarray}
\bscE &=& - \alpha_1^{(\Omega)} (\bg \cdot \bOmega) \bmB
    - \alpha_2^{(\Omega)} ((\bOmega \cdot \bmB) \bg + (\bg \cdot \bmB) \bOmega)
\nonumber\\
&&  - \alpha_1^{(W)} (\bg \cdot \bW) \bmB
    - \alpha_2^{(W)} ((\bW \cdot \bmB) \bg + (\bg \cdot \bmB) \bW)
\nonumber\\
&&  - \alpha^{(D)} \hat{\balpha}(\bg, \bD) \circ \bmB
\nonumber\\
&&  - (\gamma^{(0)} \bg
    + \gamma^{(\Omega)} \bg \x \bOmega
    + \gamma^{(W)} \bg \x \bW
    + \gamma^{(D)} \bg \circ \bD ) \x \bmB
\nonumber\\
&&  - \beta^{(0)} \bnab \x \bmB
    - \beta^{(D)} \bD \circ (\bnab \x \bmB)
\nonumber\\
&&  - (\delta^{(\Omega)} \bOmega + \delta^{(W)} \bW) \x (\bnab \x \bmB)
\label{eq27}\\
&&  - (\kappa^{(\Omega)} \bOmega + \kappa^{(W)} \bW) \circ (\bnab \bmB)^{(s)}
    - \kappa^{(D)} \, \hat{\bkappa} (\bD) \circ (\bnab \bmB)^{(s)} \, ,
\nonumber
\end{eqnarray}
where $\hat{\balpha}(\bg, \bD)$ is a symmetric tensor defined by
${\hat{\alpha}}_{ij}
= (\epsilon_{ilm} D_{lj} + \epsilon_{jlm} D_{li}) g_m$.
In the following the induction effects described by the individual contributions
to $\bscE$ are called $\alpha^{(\Omega)}$--effect, $\alpha^{(W)}$--effect, $\cdots$
$\kappa^{(D)}$--effect or, more summarizing, $\alpha$--effects, $\beta$--effects,
$\cdots$ $\kappa$--effects.

The occurrence of an $\alpha$--effect as a consequence of the
gradient of the turbulence intensity and the Coriolis force, that
is, of $\bg$ and $\bOmega$, is known for a long time. The
possibility of an $\alpha$--effect due to a combination of $\bg$ and
$\bW$, or of $\bg$ and $\bD$, has however been revealed only
recently \citep{rogachevskiietal03,raeste05}. Likewise the
$\gamma$--effect and its modification by the Coriolis force, that is
$\bOmega$, is known for a long time, but modifications by $\bW$ and
$\bD$ have been found only in the last--mentioned investigations.

\subsection{$\alpha$--effect}
\label{alphapec}

We still remain in the co--moving frame defined above, in which $\bmU$ is given by (\ref{eq15}),
and consider a point with $r = r_0$.
As a consequence of (\ref{eq15}) both $\bW$ and $\bD$ take specific forms.
The non--zero components of $\bW$ are $W_\varphi = - \dd {\ol{U}}_z / \dd r$
and $W_z = r \dd \Omega / \dd r$.
As can be concluded with the help of the relations in the Appendix the only non--zero elements of $\bD$
are $D_{r \varphi} = D_{\varphi r} = \frac{r}{2} \, \dd \Omega / \dd r$
and $D_{r z} = D_{z r} = \frac{1}{2} \, \dd {\ol{U}}_z / \dd r$.
As for $\bOmega$ we refer to (\ref{eq17}).

In view of the Perm dynamo experiment the simplest assumption on $\bg$
is that it is a radial vector.
In this case the tensor $\balpha$ has a remarkable peculiarity.
The terms with $\alpha_1^{(\Omega)}$ and  $\alpha_1^{(W)}$ disappear.
In the cylindrical co--ordinate system $\balpha$ has the form
\begin{displaymath}
\balpha = \left(
   \begin{array}{ccc}
   0 & \alpha_1 & \alpha_2 \\
   \alpha_1 & 0 & 0 \\
   \alpha_2 & 0 & 0 \\
   \end{array}
   \right)
\end{displaymath}
\begin{equation}
\alpha_1 = -(\alpha_2^{(W)} - \frac{1}{2} \alpha^{(D)})\, g \, \frac{\dd {\ol{U}}_z}{\dd r} \, , \quad
\alpha_2 = (\alpha_2^{(W)} - \frac{1}{2} \alpha^{(D)})\, g \, \frac{\dd {\ol{U}}_\varphi}{\dd r}
   + \alpha_2^{(\Omega)} \, g \, \Omega \, ,
\label{eq31}
\end{equation}
where ${\ol{U}}_\varphi$ and ${\ol{U}}_z$ are, of course,
measured in the rotating frame of reference.
All diagonal elements of $\balpha$ are equal to zero.
Therefore a mean magnetic field in $z$ direction does not generate
any mean electromotive force in $\varphi$ or $z$ direction.
The last statement remains true even if we relax the assumption on linearity
with respect to $\bg$.

In this context experiments with torus--shaped devices similar to
but smaller than that prepared for the dynamo experiment are of
interest. In an experiment with water instead of sodium clearly
helical small-scale motions have been observed \citep{fricketal02}.
This suggests the existence of an $\alpha$--effect in
the conducting fluid. In the case of a complete $\alpha$--effect an
azimuthal mean magnetic field leads to a mean electromotive force
with a non--zero azimuthal component. However, our above
result concerning $\balpha$, with the $z$ direction interpreted as
the azimuthal one, does not predict such a component. The existence
of helical motions alone is not sufficient for the occurrence of a
complete $\alpha$--effect. Another experiment with gallium instead
of sodium was carried out, too \citep{noskovetal04}. It clearly
confirmed that there is indeed no azimuthal electromotive force or
electric current due to an azimuthal magnetic field.

\section{Specific results for the mean electromotive force $\bscE$}
\label{results}

Our considerations on the structure of the electromotive force
$\bscE$ did not provide us with relations between the coefficients
$\alpha^{(\Omega)}_1$, $\alpha^{(\Omega)}_2$, $\cdots$,
$\kappa^{(D)}$ and the parameters of the turbulent flow. We rely
here on recent calculations of these coefficients presented in
detail in a paper by \citet{raeste05}. They were carried out just
under the restrictions introduced above, namely that $\bscE$ is only
weakly influenced by the Coriolis force and the gradient of the mean
velocity so that it is linear in $\bOmega$, $\bW$ and $\bD$. In
addition, again in the sense of restrictions already made, it was
assumed that all mean quantities determining $\bscE$ vary only
weakly in space and time so that it is also linear in the operator
$\bnab$ acting on quantities like $\ol{{\bu}^2}$ or $\bmB$, and
contains no time derivatives of them. Finally the second--order
correlation approximation was used. In this approximation the velocity
correlation tensor $v_{ij}$ of the second
rank,
\begin{equation}
v_{ij} (\bR, T; \br, t)
   = \ol{u_i (\bR + \br/2, T + t/2) u_j (\bR - \br/2, T - t/2)} \, ,
\label{eq41}
\end{equation}
plays a central part.
The calculations have partially been carried out in the Fourier space,
and the Fourier--transformed correlation tensor ${\tilde{v}}_{ij}$ has been used,
defined such that
\begin{equation}
v_{ij} (\bR, T; \br, t)
   = \int \!\!\! \int {\tilde{v}}_{ij} (\bR, T; \bk, \omega) \,
   \exp(\iu (\bk \cdot \br) - \omega t) \, \dd^3 k \, \dd \omega \, .
\label{eq43}
\end{equation}
The ``original" turbulence, which occurs in the limit of vanishing Coriolis force
and vanishing gradient of the mean motion, is assumed to deviate
from a homogeneous isotropic and mirror--symmetric one
only by a gradient of its intensity.
The corresponding Fourier--transformed correlation tensor ${\tilde{v}}_{ij}$
was assumed to have the form
\begin{equation}
{\tilde{v}}_{ij} (\bR, T; \bk, \omega)
   = \frac{1}{2}\big( \delta_{ij} - \frac{k_i k_j}{k^2}
   + \frac{\iu}{2 k^2}(k_i \nabla_j - k_j \nabla_i) \big) \, W(\bR, T; k, \omega) \, ,
\label{eq45}
\end{equation}
with some function $W$ describing the kinetic energy distribution in the Fourier space.
It was chosen such that
\begin{equation}
\ol{\bu (\bR + \br/2, T + t/2) \cdot \bu (\bR - \br/2, T - t/2)}
  = \ol{u^2} (\bR, T) \exp( - r^2 / 2 \lambda^2_c - t / |\tau_c|) \, ,
\label{eq46}
\end{equation}
where $\ol{u^2}$ describes the turbulence intensity in the limit
of vanishing Coriolis force and mean velocity gradient,
and $\lambda_c$ and $\tau_c$ are correlation length and time in this limit.

The results for the coefficients
$\alpha^{(\Omega)}_1$, $\alpha^{(\Omega)}_2$, $\cdots$, $\kappa^{(D)}$
can be represented with the help of the Prandtl number $P_m$ of the fluid,
the magnetic Reynolds number $R_u$ of the turbulent motion
and a dimensionless parameter $q$,
\begin{equation}
P_m = \nu / \eta \, , \quad
R_u = u \lambda_c / \eta \, , \quad
q = \lambda^2_c / \eta \tau_c  \, ,
\label{eq261}
\end{equation}
where $u = \sqrt{\ol{u^2}}$.
Clearly $R_u$, like $u$, in general depends on position.
The quantity $q$ is the ratio of the magnetic diffusion time $\lambda^2_c / \eta$
to the correlation time $\tau_c$.
We speak of low--conductivity limit if $q \to 0$,
and of high--conductivity limit if $q \to \infty$.

As for the general results concerning $\alpha^{(\Omega)}_1$, $\alpha^{(\Omega)}_2$, $\cdots$, $\kappa^{(D)}$
we refer to the mentioned paper, in particular relations (55)--(58) and figures~1 and 2.
These results cover arbitrary $P_m$ and $q$.
We may write  $\alpha^{(\Omega)}_1$, $\alpha^{(\Omega)}_2$, $\cdots$, $\kappa^{(D)}$
except $\gamma^{(0)}$ and $\beta^{(0)}$ in the form
\begin{equation}
\alpha^{(\Omega)}_1 = \tilde{\alpha}^{(\Omega)}_1 (P_m, q) \, R_u^2 \, \lambda_c^2  \, , \;
   \alpha^{(\Omega)}_2 = \tilde{\alpha}^{(\Omega)}_2 (P_m, q) \, R_u^2 \, \lambda_c^2  \, , \cdots \, ,
   \kappa^{(D)} = \tilde{\kappa}^{(D)} (P_m, q) \, R_u^2 \, \lambda_c^2 \, ,
\label{eq267}
\end{equation}
and
\begin{equation}
\gamma^{(0)} = \tilde{\gamma}^{(0)}(q) \, R_u^2 \, \eta  \, , \quad
   \beta^{(0)} = \tilde{\beta}^{(0)}(q) \, R_u^2 \, \eta  \, .
\label{eq269}
\end{equation}
The $\tilde{\alpha}^{(\Omega)}_1$, $\tilde{\alpha}^{(\Omega)}_2$, $\cdots$, $\tilde{\kappa}^{(D)}$
are dimensionless functions of $P_m$ and $q$.
Only $\tilde{\gamma}^{(0)}$, $\tilde{\beta}^{(0)}$ and $\tilde{\delta}^{(W)}$ are independent of $P_m$.
In the case of $\tilde{\gamma}^{(0)}$ and $\tilde{\beta}^{(0)}$ this independence appears as natural
when considering the derivation of these results,
in the case of $\tilde{\delta}^{(W)}$ it occurs as a accidental compensation of different influences.
Further we have $\tilde{\gamma}^{(0)} = \frac{1}{2} \, \tilde{\beta}^{(0)}$
and $\tilde{\gamma}^{(\Omega)} = \tilde{\delta}^{(\Omega)}$.
As will become clear in the following section here only results for specific $P_m$ and $q$
are of interest.
Table~1 gives some values of the $\tilde{\alpha}^{(\Omega)}_1$, $\tilde{\alpha}^{(\Omega)}_2$, $\cdots$,
$\tilde{\kappa}^{(D)}$ for such $P_m$ and $q$.

\begin{table}
\tbl{Numerical values of $\tilde{\alpha}^{(\Omega)}_1$,
$\tilde{\alpha}^{(\Omega)}_2$, $\cdots$, $\tilde{\kappa}^{(D)}$ for
various values of $P_m$ and $q$.}
{\begin{tabular}{|l|l|l|l|l|l|l|l|l|l|} \hline
$P_m$ & 1 & $10^{-1}$ & $10^{-2}$ & $10^{-3}$ & $10^{-4}$ & $10^{-5}$ & $10^{-5}$ & $10^{-5}$ & $10^{-6}$ \\
$q$ & 0.02 & $0.02$ & $0.02$ & $0.02$ & $0.02$ & $0.002$ & $0.02$ & $0.2$ & $0.02$ \\ \hline
$\tilde{\alpha} _{1}^{(\Omega )}$ & 0.0792 & 0.689 & 3.47 & 7.5 & 8.84 & 0.982 & 9.01 & 87.1 & 9.03 \\
$\tilde{\alpha} _{2}^{(\Omega )}$ & -0.104 & -0.654 & -1.98 & -2.35 & -2.2 & -0.195 & -2.18 & -22.5 & -2.17 \\
$\tilde{\alpha} _{1}^{(W)}$ & 0.0439 & 0.38 & 1.82 & 3.76 & 4.38 & 0.46 & 4.46 & 43.6 & 4.47 \\
$\tilde{\alpha} _{2}^{(W)}$ & -0.0094 & -0.139 & -0.57 & -0.972 & -1.07 & -0.092 & -1.08 & -10.9 & -1.08 \\
${{\alpha }^{(D)}}$ & -0.0437 & -0.171 & -0.374 & -0.191 & -0.048 & -0.015 & -0.027 & -0.266 & -0.025 \\
$\tilde{\gamma }^{(0)}$ & 0.0546 & 0.0546 & 0.0546 & 0.0546 & 0.0546 & 0.049 & 0.0546 & 0.0555 & 0.0546 \\
$\tilde{\gamma}^{(\Omega )}$ & -0.004 & -0.028 & -0.067 & -0.089 & -0.093 & -0.066 & -0.093 & -0.105 & -0.094 \\
$\tilde{\gamma}^{(W)}$ & -0.0086 & -0.0541 & -0.409 & -1.12 & -1.39 & -0.165 & -1.42 & -13.5 & -1.43 \\
$\tilde{\gamma}^{(D)}$ & -0.0888 & -0.766 & -3.09 & -5.44 &  -6.05 & -0.629 & -6.13 & -60.3 & -6.13 \\
$\tilde{\beta}^{(0)} $& 0.109 & 0.109 & 0.109 & 0.109 & 0.109 & 0.0979 & 0.109 & 0.111 & 0.109 \\
$\tilde{\beta}^{(D)}$ & 0.059 & 0.558 & 2.03 & 3.09 & 3.25 & 0.293 & 3.27 & 33.1 & 3.27 \\
$\tilde{\delta}^{(\Omega )}$ & -0.004 & -0.028 & -0.067 & -0.089 & -0.093 & -0.066 & -0.093 & -0.105 & -0.094 \\
$\tilde{\delta}^{(W)}$ & 0.0214 & 0.0214 & 0.0214 & 0.0214 & 0.0214 & 0.0125 & 0.0214 & 0.0255 & 0.0214 \\
$\tilde{\kappa}^{(\Omega )}$ & 0.0072 & 0.098 & 0.268 & 0.364 & 0.382 & 0.24 & 0.386 & 0.25 & 0.386 \\
$\tilde{\kappa}^{(W)}$ & -0.0052 & 0.0468 & 0.14 & 0.19 & 0.202 & 0.13 & 0.204 & 0.236 & 0.204 \\
$\tilde{\kappa }^{(D)}$ & 0.114 & 0.631 & 2.13 & 3.21 & 3.38 & 0.373 & 3.39 & 33.2 & 3.4 \\
\hline
\end{tabular}}
\end{table}

\section{The effect of turbulence in the Perm dynamo device}
\label{estimates}

\subsection{Estimates of the turbulence effects}

Let us make a few estimates of the turbulence effects to be
expected in the Perm dynamo device on the basis of the general form (\ref{eq27})
of the mean electromotive force $\bscE$, the relations (\ref{eq267}) and (\ref{eq269}),
and numerical values represented in table~1.
As for the properties of the liquid sodium we put $\eta = 0.1 \mbox{m}^2 / \mbox{s}$ and $P_m = 10^{-5}$.
As a typical value $U$ of the mean velocity $\bmU$ we
use $U = 50 \mbox{m} / \mbox{s}$. For the velocity $u$ of the
turbulent motion, the correlation length $\lambda_c$ and the
correlation time $\tau_c$ we assume $u = 1 \mbox{m} / \mbox{s}$,
$\lambda_c = 0.01 \, \mbox{m}$ and $\tau_c = 0.05 \, \mbox{s}$.
This implies $R_u = 0.1$ and $q = 0.02$.
Further $|\bg|$ is estimated by $1/R$, where $R$ means the radius of the fluid
cylinder. Likewise $|\bOmega|$, $|\bW|$ and $|\bD|$ are estimated
by $U/R$, and spatial derivatives of $\bmB$ by $\bmB/R$.
We put $R = 0.12 \mbox{m}$.

Relation (\ref{eq27}) for $\bscE$ ignores any time variation of $\bmB$.
Since we are in the limit of small $q$ this is justified as long as the characteristic time
of this variation is large compared to $\lambda^2_c / \eta$, with the above data $10^{-3} \mbox{s}$.
The calculations described below show that this characteristic time is not smaller than $10^{-2}$
so that this condition is satisfied.

Consider first the $\beta$--effects, which may be interpreted by introducing
a mean--field version of the magnetic diffusivity instead of the molecular diffusivity $\eta$.
This mean--field version is in general a tensor,
$\eta_{\mathrm{m} \, {\it ij}} = (\eta + \beta^{(0)}) \delta_{ij} + \beta^{(D)} D_{ij}$.
The contribution of the $\beta^{(0)}$--effect to the mean--field diffusivity,
measured in units of $\eta$, is $\beta^{(0)}/\eta = \tilde{\beta}^{(0)} R^2_u $.
With the above data it takes a value of about $10^{-3}$.
The corresponding contribution of the $\beta^{(D)}$--effect is of the order of
$\beta^{(D)} U / \eta R = \tilde{\beta}^{(D)} (U R / \eta) \, (\lambda_c / R)^2 R^2_u$,
with the above data about $1.3 \cdot 10^{-2}$.

It is natural to compare the $\gamma$--effects, which describe a transport of mean magnetic flux,
with that of the mean velocity $\bmU$.
As for the $\gamma^{(0)}$--effect we find that $|\gamma^{(0)} \bg| / |\bmU|$
is of the order of $\gamma^{(0)} / U R = \tilde{\gamma}^{(0)}(\eta / U R) \, R^2_u$,
which takes a value of about $10^{-5}$.
In analogous estimates for the $\gamma^{(\Omega)}$, $\gamma^{(W)}$ and $\gamma^{(D)}$--effects
the ratios corresponding to $\gamma^{(0)} / U R$ are $\gamma^{(\Omega)}/ R^2$,
$\gamma^{(W)} / R^2$ and $\gamma^{(D)} / R^2$, all being independent of $U$.
The first two are smaller than the last one, which is about $4 \cdot 10^{-4}$.

We may also compare the $\alpha$, $\delta$ and $\kappa$--effects with that of the mean velocity $\bmU$.
They are then characterized by the ratios
$\alpha_1^{(\Omega)} / R^2$, $\alpha_2^{(\Omega)} / R^2$, $\alpha_1^{(W)} / R^2$, $\cdots$,
$\delta^{(\Omega)}/ R^2$, $\cdots$, $\kappa^{(\Omega)} /R^2$, $\cdots$,
again being independent of $U$.
As for the $\alpha_1^{(\Omega)} / R^2$, $\alpha_2^{(\Omega)} / R^2$, $\cdots$ $\alpha^{(D)} / R^2$,
the first one is about $6 \cdot 10^{-4}$ but all others are smaller.
We recall that the $\alpha_1^{(\Omega)}$ or the $\alpha_1^{(W)}$--effects vanish
if $\bg$ is orthogonal to $\bOmega$ or $\bW$, respectively.
Further $\delta^{(\Omega)}/ R^2$ is about $7 \cdot 10^{-6}$ but $\delta^{(W)}/ R^2$ smaller.
Among the $\kappa^{(\Omega)}$, $\cdots$ $\kappa^{(D)} /R^2$ the last one is about $2 \cdot 10^{-4}$
but the others are smaller.

We may conclude from these estimates that the influence of turbulence on the screw dynamo
in the Perm experiment should be rather small.
There is hardly the possibility of another type of dynamo,
which would become possible by turbulence effects.

\subsection{Dynamo model}

Some more detailed numerical investigations concerning the effects of turbulence
on the screw dynamo have been carried out,
mainly with the intention to find out in which sense they influence its excitation condition.
Again $\bg$ was considered as a radial vector such that $\bnab \ol{u^2} = \bg \, \ol{u^2}$.

Let us first give equation (\ref{eq05}) governing $\bmB$ inside the cylinder and the supplementing relations
in a non--dimensional form.
In that sense we put
\begin{equation}
\Omega = \Omega_0 \tilde{\Omega} \, , \quad \ol{U}_z = U_0 \tilde{U}_z \, , \quad
    u^2 = u^2_0 \chi \, .
\label{eq299}
\end{equation}
Here $\Omega_0$ and $U_0$ are constants describing typical values of $\Omega$ and $\ol{U}_z$,
e.g., being their values at $r = 0$,
and $\tilde{\Omega}$ and $\tilde{\ol{U}}_z$ are dimensionless functions of $r$.
Further $u_0$ means the value of $u$ at $r = 0$, and $\chi$ is again a dimensionless function of $r$.
In the following we measure all lengths in units of the radius $R$ of the cylinder
and the time in units of $R^2 / \eta$.
Then (\ref{eq05}) can be written in the form
\begin{equation}
\partial_t \bmB - \bnab \x \big( R_U \, \tilde{\bU} \x \bmB
   + R^2_{u0} \, (\tilde{\bF}(\bmB) + \xi R_U \tilde{\bG} (\bmB) \, ) \big)
   - \bnab^2 \bmB = \bzo \, , \quad
   \bnab \cdot \bmB = 0
\label{eq301}
\end{equation}
with the parameters
\begin{equation}
R_U = \sqrt{\Omega_0^2 R^2 + U_0^2} \, R / \eta  \, , \quad R_{u0} = u_0 \lambda_c / \eta \, , \quad
    \xi = (\lambda_c / R)^2 \, .
\label{eq311}
\end{equation}
and
\begin{equation}
\tilde{\bU} = \big( 0 \, , \zeta_\Omega \, \tilde{\Omega} (r) \, r \, , \zeta_U \, \tilde{U}_z \, \big)
\label{eq305}
\end{equation}
\begin{equation}
\tilde{\bF} =  - \tilde{\gamma}^{(0)} \, \chi' \, \hat{\br} \x \bmB
    - \tilde{\beta}^{(0)} \, \chi \, \bnab \x \bmB
\label{eq307}
\end{equation}
\begin{eqnarray}
\tilde{\bG} &=& - \tilde{\alpha}_2^{(\Omega)} \, \chi' \, \zeta_\Omega \, \tilde{\Omega} \,
    \big(\mB_z \hat{\br} + \mB_r \hat{\bz} \big)
\nonumber\\
&& - (\tilde{\alpha}_2^{(W)} - \tilde{\alpha}^{(D)}) \, \chi' \, \big( \zeta_\Omega \, r \tilde{\Omega}' \,
    (\mB_z \hat{\br} + \mB_r \hat{\bz})
    - \zeta_U \, \tilde{U}_z^{\prime} \,
    (\mB_\varphi \hat{\br} + \mB_r \hat{\bvarphi}) \big)
\nonumber\\
&& - \big( \tilde{\gamma}^{(\Omega)} \chi' \, \zeta_\Omega \, \tilde{\Omega} \hat{\bvarphi}
     + (\tilde{\gamma}^{(W)} - \tilde{\gamma}^{(D)}) \, \chi' \,
     (\zeta_\Omega \, r \tilde{\Omega}' \hat{\bvarphi} + \zeta_U \, \tilde{U}_z^{\prime} \hat{\bz}) \big) \x \bmB
\nonumber\\
&& - \tilde{\beta}^{(D)} \chi \, \big(\zeta_\Omega \, r \, \tilde{\Omega}' \,
    ((\bnab \x \bmB)_\varphi \hat{\br} + (\bnab \x \bmB)_r \hat{\bvarphi})
    + \zeta_U \, \tilde{U}_z^{\prime} \,
    ((\bnab \x \bmB)_z \hat{\br} + (\bnab \x \bmB)_r \hat{\bz}) \big)
\nonumber\\
&& - \big( \tilde{\delta}^{(\Omega)} \, \chi \, \zeta_\Omega \, \tilde{\Omega} \, \hat{\bz}
    + \tilde{\delta}^{(W)} \, \chi \, (\zeta_\Omega \, r \, \tilde{\Omega}' \, \hat{\bz}
    - \zeta_U \, \tilde{U}_z^{\prime} \, \hat{\bvarphi}) \big)
    \times (\bnab \times \bmB)
\label{eq309}\\
&& + \big(\tilde{\kappa}^{(\Omega)} \, \chi \, \zeta_\Omega \, \tilde{\Omega} \, \hat{\bz}
    - \tilde{\kappa}^{(W)} \, \chi \, (\zeta_\Omega \, r \, \tilde{\Omega}' \, \hat{\bz}
    - \zeta_U \, \tilde{U}_z^{\prime} \, \hat{\bvarphi}) \big)
    \circ (\bnab \bmB)^{(s)}
\nonumber\\
&&  + \tilde{\kappa}^{(D)} \, \chi \, \big(\zeta_\Omega \, r \, \tilde{\Omega}' \, ((\bnab \bmB)^{(s)}_{rz} \hat{\br}
    - (\bnab \bmB)^{(s)}_{\varphi z} \hat{\bvarphi}
    - ((\bnab \bmB)^{(s)}_{rr} - (\bnab \bmB)^{(s)}_{\varphi \varphi})\hat{\bz})
\nonumber\\
&& \qquad \qquad - \zeta_U \, \tilde{U}_z^{\prime} \, ((\bnab \bmB)^{(s)}_{r \varphi} \hat{\br}
    - ((\bnab \bmB)^{(s)}_{rr} - (\bnab \bmB)^{(s)}_{zz}) \hat{\bvarphi}
    - (\bnab \bmB)^{(s)}_{\varphi z}\hat{\bz}) \big)
\nonumber
\end{eqnarray}
\begin{equation}
\zeta_\Omega = 1 / \sqrt{1 + \zeta^2} \, , \quad
    \zeta_U = \zeta / \sqrt{1 + \zeta^2} \, , \quad
    \zeta = U_0 / \Omega_0 R \, .
\label{eq313}
\end{equation}
The $\hat{\br}$, $\hat{\bvarphi}$ and $\hat{\bz}$ are unit vectors in $r$, $\varphi$ and $z$ direction,
and primes indicate differentiations with respect to the dimensionless radial co--ordinate $r$.
Note that the $\tilde{\gamma}^{(0)}$, $\tilde{\beta}^{(0)}$, $\tilde{\alpha}_2^{(\Omega)}$, $\cdots$
$\tilde{\kappa}^{(D)}$ depend on $q$, most of them also on $Pm$.
As for the components of $(\bnab \bmB)^{(s)}$ in the cylindrical co--ordinate system
we refer to the Appendix.
Of course, equation (\ref{eq301}) has to be completed by boundary conditions.

We will look for solutions of the above equations for $\bmB$ having the form
\begin{equation}
\bmB = \hat{\bmB} \, \exp(\iu(m \varphi + k z) + \lambda t)
\label{eq321}
\end{equation}
with a positive integer $m$, a real non--zero wavenumber $k$ and a complex growth rate $\lambda$.
Note that by Cowling's theorem solutions with $m = 0$ have to be excluded as long as $R_{u0} = 0$.

\begin{figure}
 \includegraphics[width=0.32\textwidth]{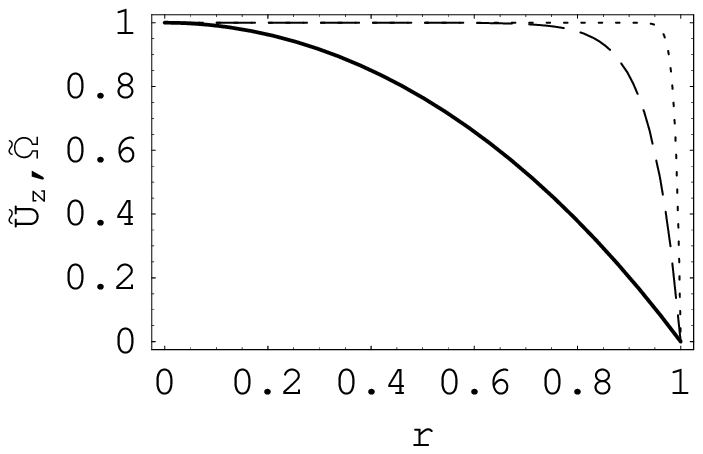}
 \includegraphics[width=0.32\textwidth]{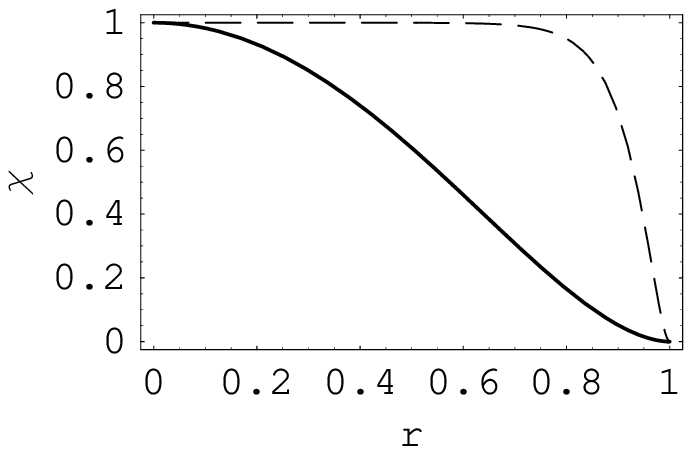}
 \includegraphics[width=0.32\textwidth]{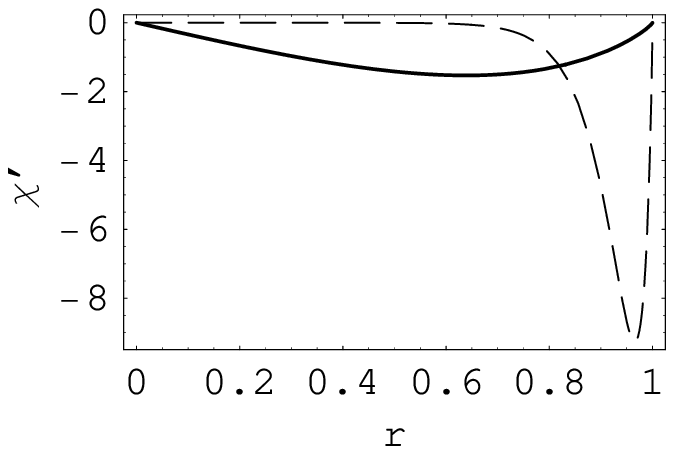}
 \caption{Profiles of $\tilde{\Omega}$ and $\tilde{U}_z$ (left), and of $\chi$ (middle)
 and $\chi'$ (right)
 for $\rho =1$ (solid), $\rho = 18$ (dashed) and $\rho = 100$ (dotted)}
\label{profiles}
\end{figure}

As for $\tilde{\Omega}$ and $\tilde{U}_z$ we put
\begin{equation}
\tilde{\Omega} (r) = \tilde{U}_z (r) = \frac{\cosh (\rho) - \cosh (\rho r)}{\cosh (\rho) - 1}
\label{eq323}
\end{equation}
with some parameter $\rho$. Interpreted in view of $\tilde{U}_z$,
for $\rho =1$ the flow has a Poiseuille profile, for $\rho \to
\infty$ a piston profile. The outcomes of the water experiments made
to simulate the flow in the dynamo device fit best to $\rho = 18$.
In view of $\chi$, which is defined by (\ref{eq299}), we adopt a
relation for the r.m.s. value $u$ of the turbulent velocity given by
\citet{schlichting64},
\begin{equation}
u \big( 2.5 \ln {{u R}\over\nu} +1.75 \big) = U_z \, .
\label{eq325}
\end{equation}
On this basis $\chi$ has been calculated with $\nu$ such that $Pm = 10^{-5}$.
Profiles of $\tilde{\Omega}$ and  $\tilde{U}_z$ and of $\chi$ and $\chi^\prime$ for several $\rho$
are depicted in figure~\ref{profiles}.

All calculations reported in the following have been carried out with $Pm = 10^{-5}$ and $q = 0.02$.
The corresponding numerical values of $\tilde{\alpha}_2^{(\Omega)}$,
$\tilde{\alpha}_2^{(W)}$, $\cdots$, $\tilde{\kappa}^{(D)}$ are given in table~1.

As for the surroundings of the fluid cylinder two cases are considered.
In the first one all surroundings are assumed to be free space.
Since only non--zero $k$ are admitted the total electric current
through a cross--section of the cylinder is equal to zero.
Therefore it is to be required that $\bmB$ inside the cylinder
continues as a single-valued potential field in outer space.
In the second case it is assumed that the fluid cylinder is muffled
by a rigid electrically conducting shell, in the experimental device made of copper,
and there is free space outside this shell.
Again $\bmB$ has to continue as a single--valued potential field in this free space.
The thickness of the shell, measured in units of $R$, is denoted by $d$,
and the ratio of the electrical conductivity of the shell to that of the fluid by $s$.
The first case may be considered as the special case $d = 0$ of the second one.
The experimental device corresponds to $d=0.15$ and $s = 5$.

The equations (\ref{eq301})--(\ref{eq309}) together with the ansatz (\ref{eq321})
have been reduced to a system of ordinary differential equation which,
together with the boundary conditions, defines an eigenvalue problem with
the growth rate $\lambda$ as eigenvalue parameter.
After discretization of the equations this problem has been solved with the QR algorithm.
In the following we restrict our attention to solutions with $\Re (\lambda) = 0$.
The corresponding values of $R_U$ are denoted as critical values.

\subsection{Influences of the turbulence on the excitation threshold}
\label{subsec53}

\begin{figure}
 \includegraphics[width=0.49\textwidth]{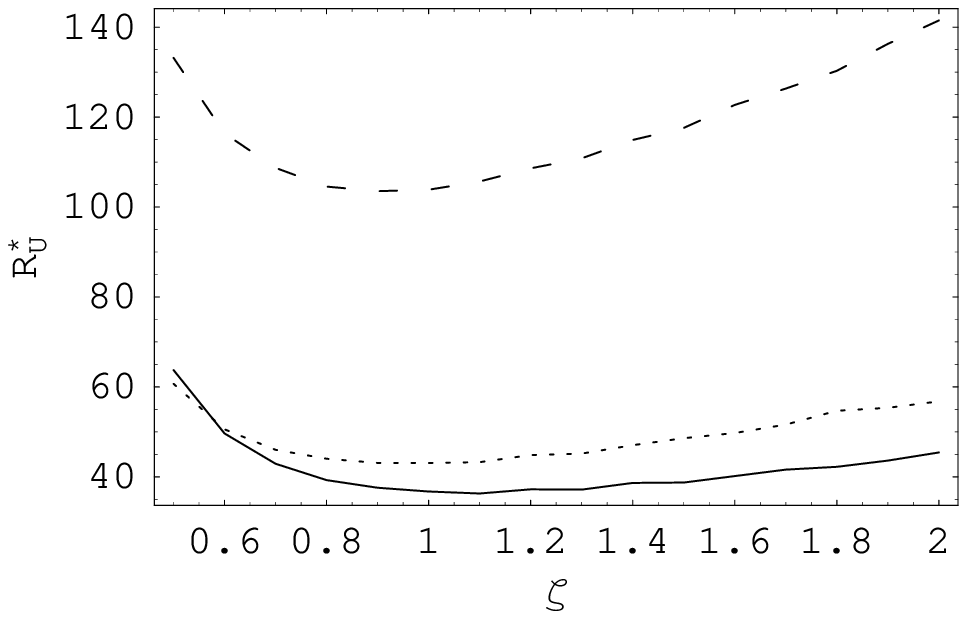}
 \includegraphics[width=0.49\textwidth]{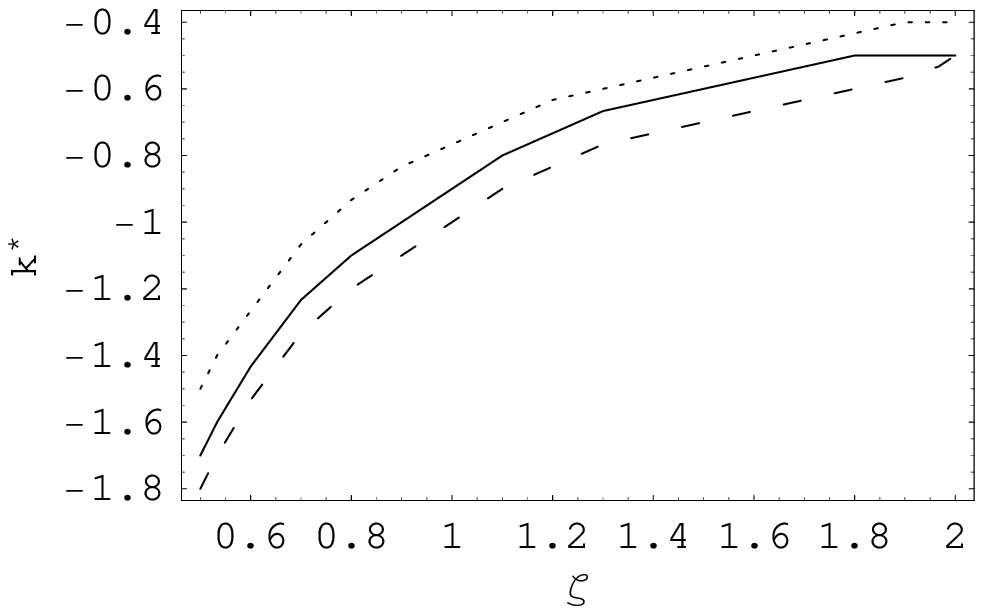}
 \caption{$R_U^*$ and $k^*$ versus $\zeta$ for the cases $d = 0$, $\rho =1$ (dashed),
 $d = 0.15$, $s = 5$, $\rho = 1$ (dotted),
 and $d = 0.15$, $s = 5$, $\rho = 18$ (solid)
 in the absence of turbulence, that is, $R_{u0} = 0$}
\label{Rcrit}
\end{figure}

Let us first give critical values of $R_U$ for the ideal situation
without turbulence, that is, $R_{u0} = 0$. For any given $\zeta$ and
$\rho$ they still depend on $m$ and $k$. We restrict ourselves to $m
= 1$. All solutions of the equations considered with the same
parameters but other $m$ have higher critical values of $R_U$. We
denote the minimal critical $R_U$ for given $\zeta$ and $\rho$ but
varying $k$ by $R_U^*$ and the corresponding $k$ by $k^*$.
Figure~\ref{Rcrit} shows $R_U^*$ and $k^*$ as functions of $\zeta$
for several cases concerning $d$ and $\rho$. It shows in particular
that a highly conducting shell around the fluid cylinder markedly
favors the screw dynamo; \citep[see
also][]{avalos-zunigaetal03,avalos-zunigaetal05}.

\begin{figure}
 \includegraphics[width=0.49\textwidth]{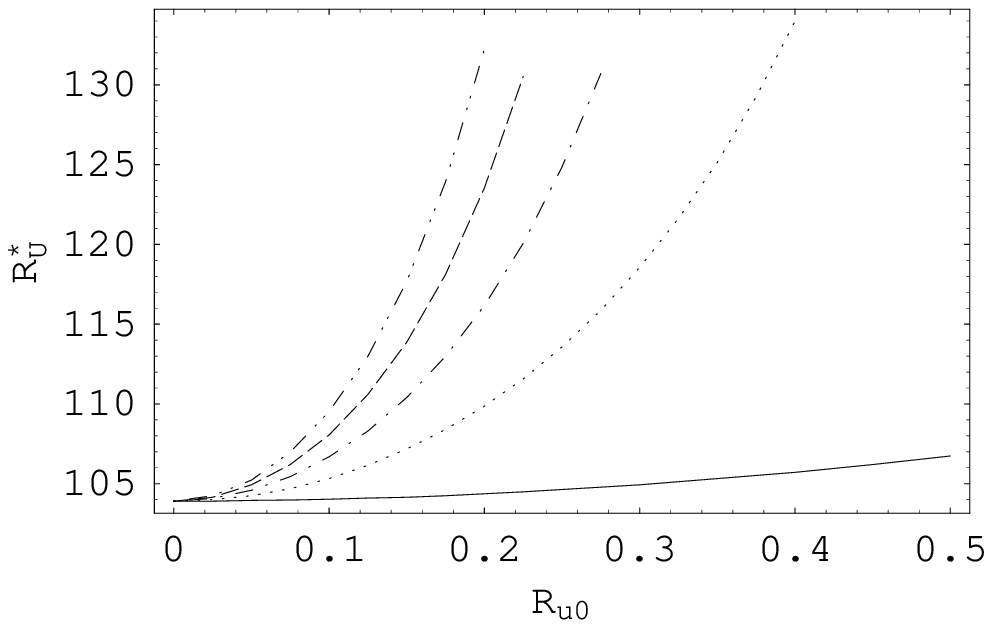}
 \includegraphics[width=0.49\textwidth]{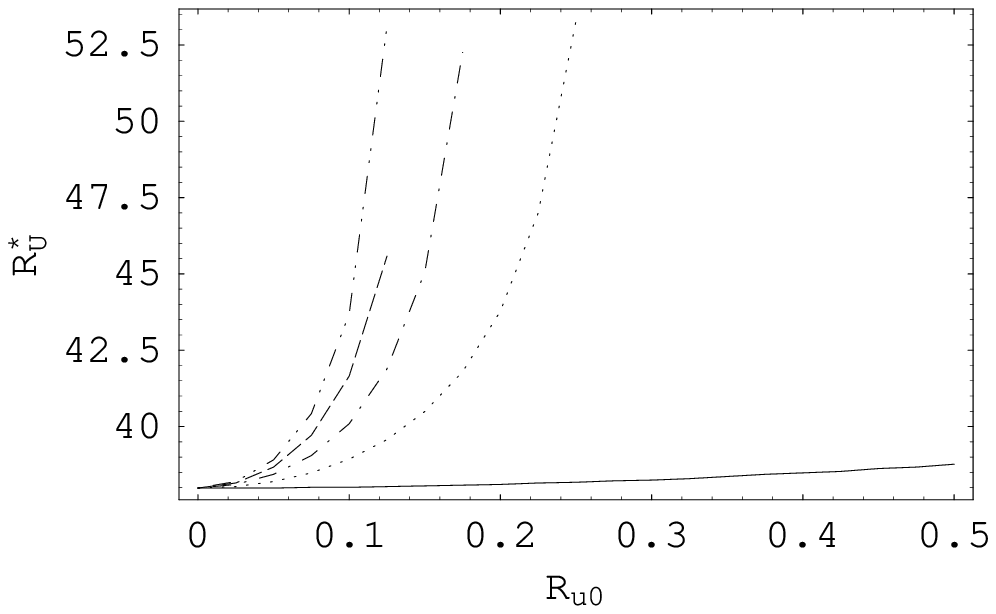}
 \hspace{-4.8cm}\raisebox{1.6cm}
 {\includegraphics[width=0.25\textwidth]{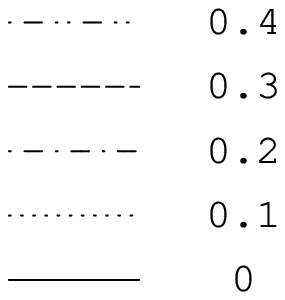}}
 \hspace{-1.5cm}\raisebox{4.5cm}{$\xi$}\hspace{.5cm}
 \hspace{-9cm}\raisebox{4.9cm}{(a)}
 \hspace{8cm}\raisebox{4.9cm}{(b)}
  \caption{$R_U^*$ versus $R_{u0}$ for $\zeta = 1$ and different $\xi$,
 with $d = 0$, $\rho = 1$ (left)
 and $d = 0.15$, $s = 5$, $\rho = 18$ (right)}
\label{allturb}
\end{figure}

We admit now turbulence.
Since in the experiment $\zeta$ will probably be close to unity we put now throughout $\zeta = 1$.
As figure~\ref{allturb} demonstrates the turbulence generally enlarges the dynamo threshold.
Consider the most realistic case $d = 0.15$, $s = 0.5$ and $\rho = 18$.
Assuming that $R_{u0}$ is of the order of $10^{-1}$ and $\xi$ of the order $10^{-2}$
we conclude that $R_U^*$ grows by at most $0.5$ percent.
We point out, however, that all our calculations have been done with $P_m =10^{-5}$ and $q = 0.02$.
The data of table~1 suggest that the influence of the turbulence will be markedly larger
if larger values of $q$ have to be assumed.

Several investigations have been made in order to find out which specific induction effects
of the turbulence such as $\alpha$--effects, $\gamma$--effects etc.
hamper the screw dynamo or possibly support it.
In that sense only parts of the mean electromotive force $\bscE$ have been taken into account.
Both cases with $\rho = 1$ and more realistic ones with $\rho = 18$ have been studied.
In the latter one parts of the profiles of $\tilde{\Omega}$, $\tilde{U}_z$ and $\chi$ are markedly steeper.

\begin{figure}
 \includegraphics[width=0.49\textwidth]{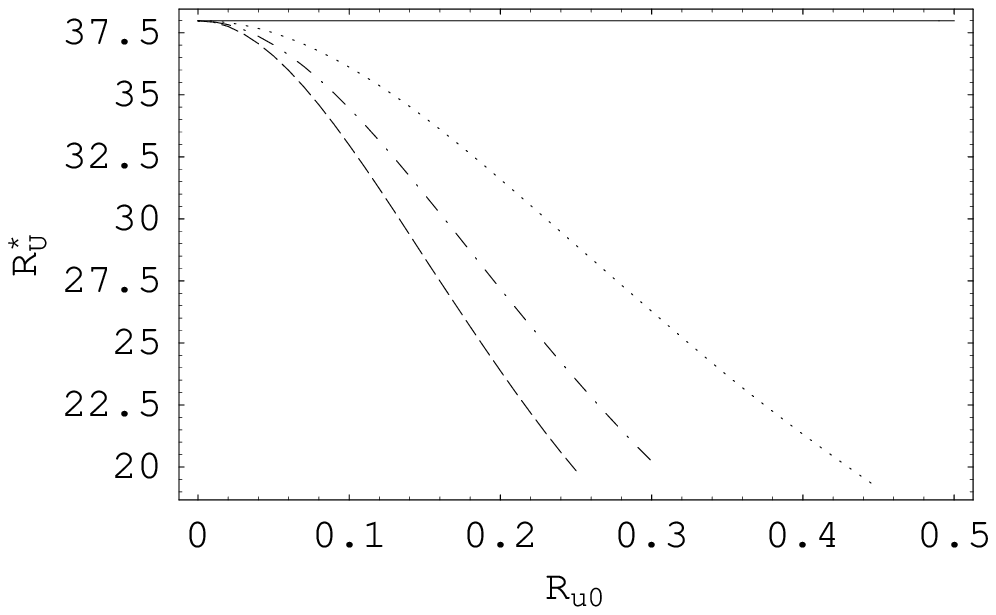}
 \includegraphics[width=0.49\textwidth]{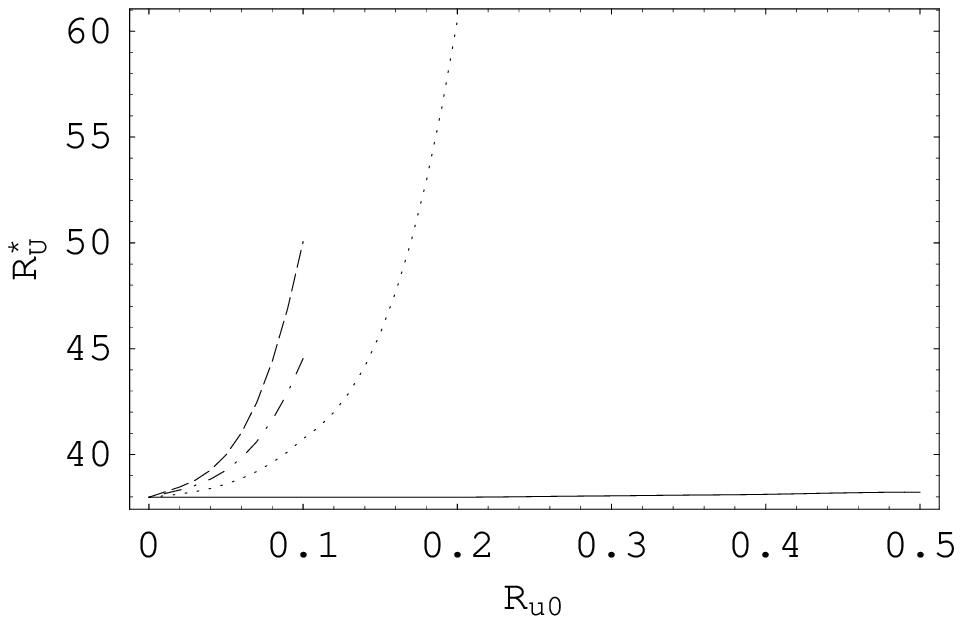}
 \hspace{-10.cm}\raisebox{4.9cm}{(a)}
 \hspace{8cm}\raisebox{4.9cm}{(b)}
 \caption{$R_U^*$ versus $R_{u0}$ for cases in which no other turbulent induction effects
 than the $\alpha$--effects (left) or the $\gamma$--effects (right) are taken into account,
 with $\zeta = 1$, $d = 0.15$, $s = 5$, $\rho =18$ and different $\xi$
 (indicated by different line styles as in figure~\ref{allturb})}
\label{alphagamma}
\end{figure}

Let us consider first the $\alpha$--effects. Note that the
$\alpha_1^{(\Omega)}$ and $\alpha_1^{(W)}$--effects are absent since
$\bg$ is orthogonal to both $\bOmega$ and $\bW$. The influence of
the $\alpha_2^{(\Omega)}$, $\alpha_2^{(W)}$ and
$\alpha^{(D)}$--effects on the screw dynamo depends crucially on
$\rho$. In cases with $\rho = 1$ it was found that they act against
the screw dynamo. Figure~\ref{alphagamma}a shows results for a more
realistic case with $\rho = 18$, in which these effects clearly
support the screw dynamo.

Consider next the $\gamma$--effects. In all investigated cases, both
with $\rho = 1$ and also with higher $\rho$, they hamper the screw
dynamo. Figure~\ref{alphagamma}b illustrates this for the above case
with $\rho = 18$.

\begin{figure}
 \includegraphics[width=0.49\textwidth]{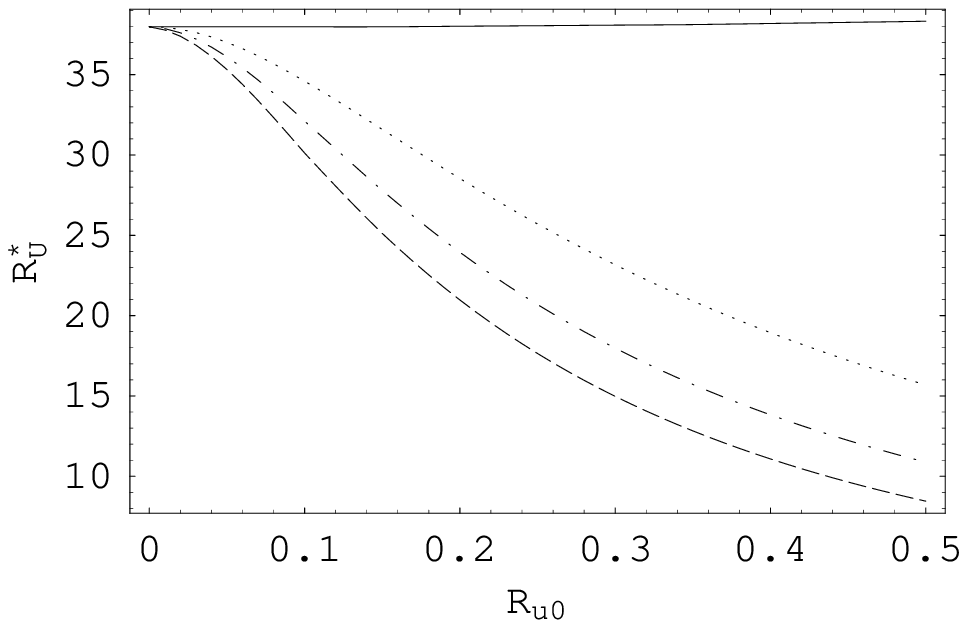}
 \includegraphics[width=0.49\textwidth]{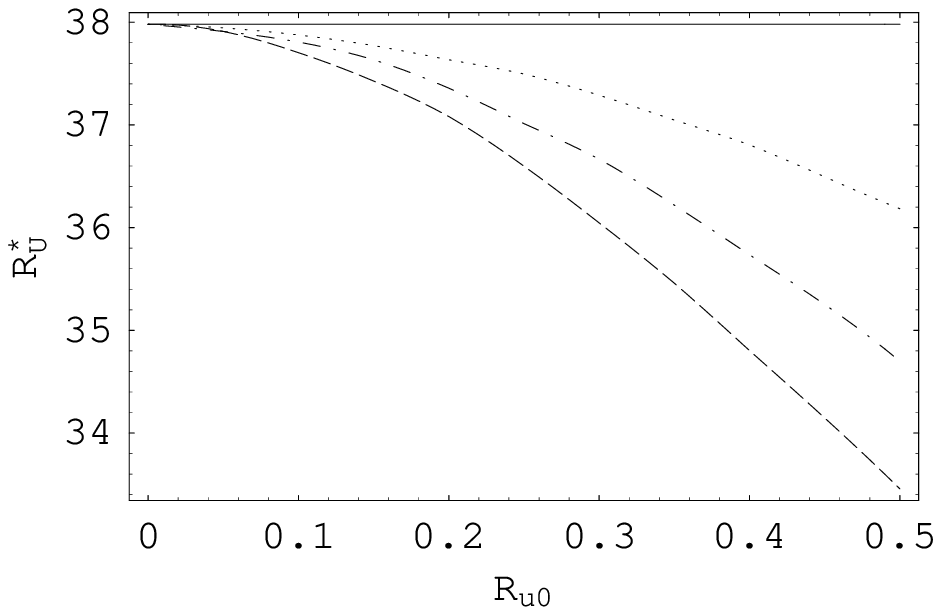}
 \hspace{-10.cm}\raisebox{4.9cm}{(a)}
 \hspace{8cm}\raisebox{4.9cm}{(b)}
 \caption{$R_U^*$ versus $R_{u0}$ for cases in which no other turbulent induction effects
 than the $\beta$--effects (left) or the $\delta$--effects (right) are taken into account,
 with $\zeta = 1$, $d = 0.15$, $s = 5$, $\rho =18$ and different $\xi$
 (indicated as in figure~\ref{allturb})}
\label{betadelta}
\end{figure}

Proceeding now to the $\beta$--effects we note first
that the $\beta^{(0)}$--effect may be interpreted in the sense of the mean--field magnetic diffusivity
$\eta + \beta^{(0)}$ different from the molecular diffusivity $\eta$.
Since $\beta^{(0)}$ is always positive it raises the threshold of the screw dynamo.
The $\beta^{(D)}$--effect makes the mean-field magnetic diffusivity anisotropic.
It is then described by the tensor $(\eta + \beta^{(0)}) \, \delta_{ij} + \beta^{(D)} D_{ij}$.
In all investigated cases, with $\rho = 1$ and also with higher $\rho$, the $\beta^{(D)}$--effect
for not too small $\xi$ always dominates the $\beta^{(0)}$--effect and clearly supports the screw dynamo.
This implies of course that the mentioned tensor is not positive definite;
otherwise it had to act against the screw dynamo.
Results for the above case with $\rho = 18$ are shown in figure~\ref{betadelta}a.

\begin{figure}
 \includegraphics[width=0.49\textwidth]{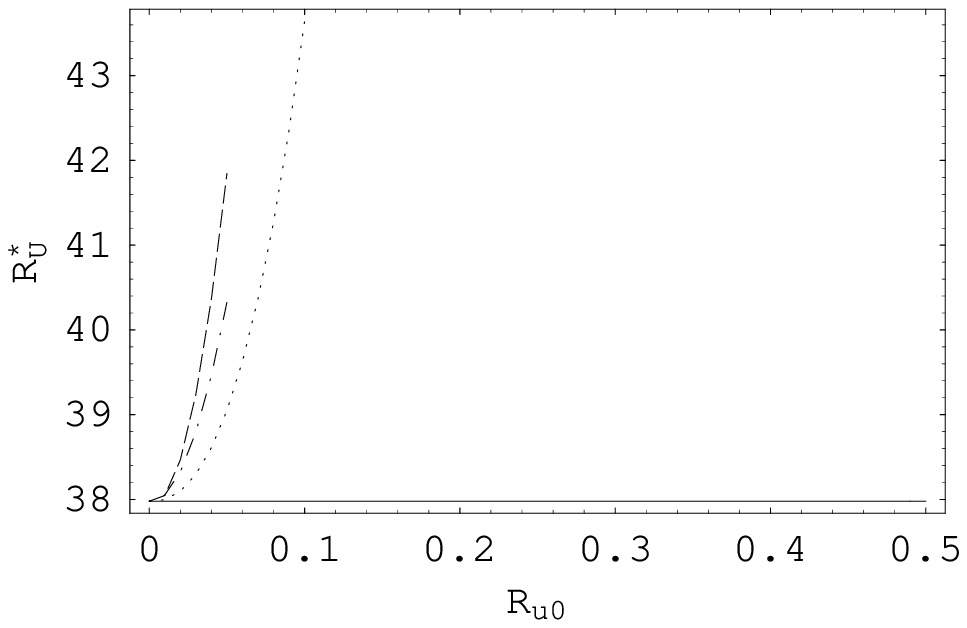}
 \includegraphics[width=0.49\textwidth]{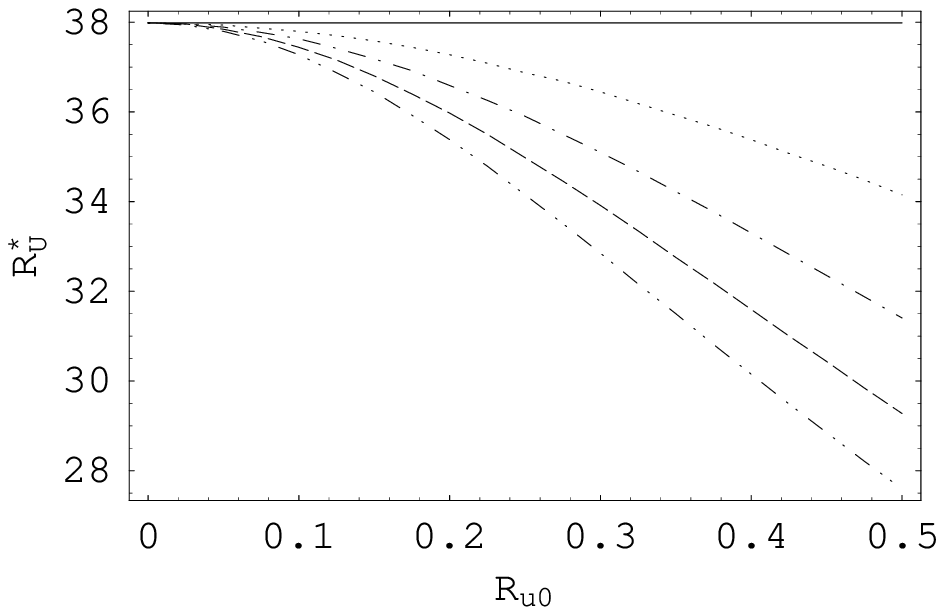}
 \hspace{-10.cm}\raisebox{4.9cm}{(a)}
 \hspace{8cm}\raisebox{4.9cm}{(b)}
 \caption{$R_U^*$ versus $R_{u0}$ for cases in which no other turbulent induction effects
 than the $\kappa$--effects (left) or the $\delta^{(\Omega)}$ and $\kappa^{(\Omega)}$--effects (right)
 are taken into account,
 with $\zeta = 1$, $d = 0.15$, $s = 5$, $\rho =18$ and different $\xi$
 (indicated as in figure~\ref{allturb})}
\label{kappaOmega}
\end{figure}

Both the $\delta^{(\Omega)}$ and $\delta^{(W)}$--effects and also the $\kappa^{(\Omega)}$--effect
always support the screw dynamo.
The $\kappa^{(W)}$ and $\kappa^{(D)}$--effects and also all $\kappa$--effects together act against it.
This is partially reflected in figures~\ref{betadelta}b and \ref{kappaOmega}a
applying again to the above case with $\rho = 18$.

We recall here that $\delta^{(\Omega)}$ and $\kappa^{(\Omega)}$--effects
together with a sufficiently strong rotational shear
open the possibility of a mean--field dynamo different from the screw dynamo,
often called $\bOmega \x \bJ$ dynamo
(R\"adler 1969, 1980, 1986, Roberts 1972, Moffatt and Proctor 1982).
The positive influence of the $\delta^{(\Omega)}$ and $\kappa^{(\Omega)}$--effects on the screw dynamo,
demonstrated by figure~\ref{kappaOmega}b, can possibly be interpreted in the sense of this dynamo mechanism.

\begin{figure}
 \includegraphics[width=0.49\textwidth]{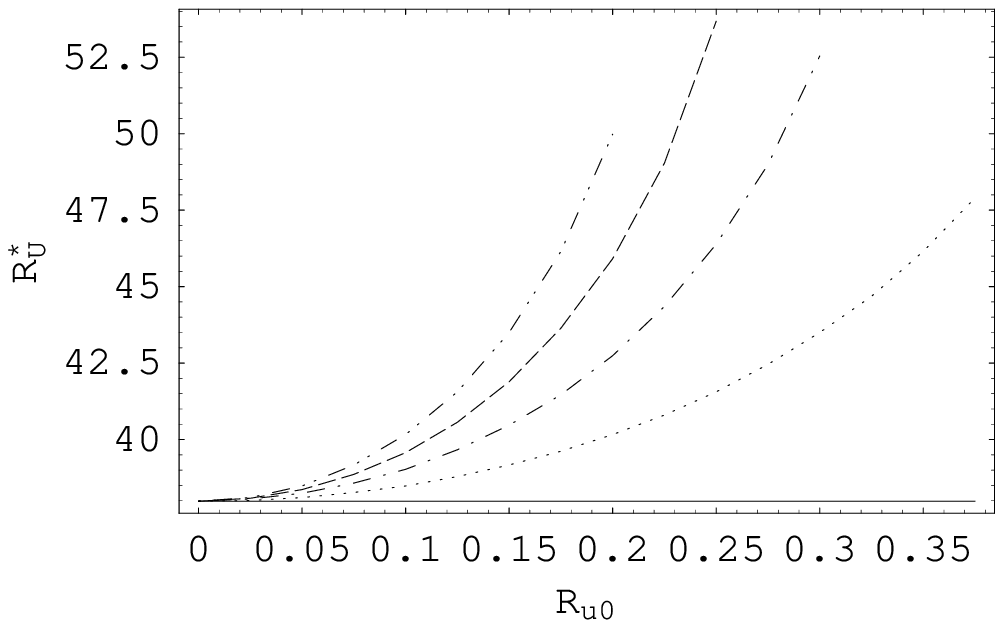}
 \includegraphics[width=0.49\textwidth]{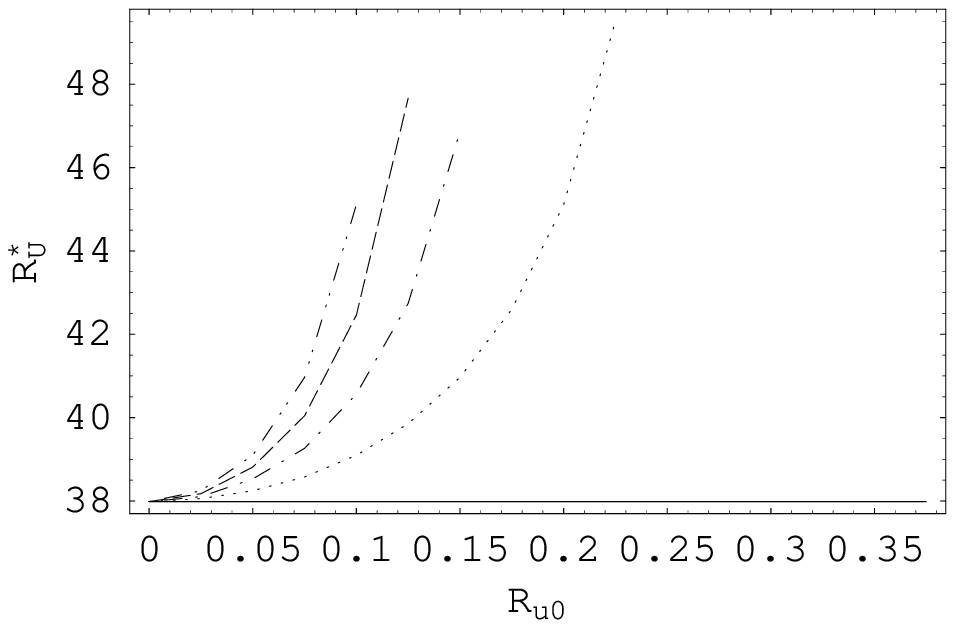}
 \hspace{-10.cm}\raisebox{4.9cm}{(a)}
 \hspace{8cm}\raisebox{4.9cm}{(b)}
 \caption{$R_U^*$ versus $R_{u0}$ for cases in which no other turbulent induction effects
 than the $\delta^{(W)}$ and $\kappa^{(W)}$--effects  (left)
 or the $\delta^{(W)}$, $\kappa^{(W)}$, $\beta^{(D)}$ and $\kappa^{(D)}$--effects (right)
 are taken into account, with $\zeta = 1$, $d = 0.15$, $s = 5$, $\rho =18$,
 and different $\xi$
 (indicated as in figure~\ref{allturb})}
\label{WandD}
\end{figure}

In contrast to the combination of the $\delta^{(\Omega)}$ and $\kappa^{(\Omega)}$--effects
that of the $\delta^{(W)}$ and $\kappa^{(W)}$--effects, as figure~\ref{WandD}a shows,
does not support the screw dynamo. For the $\bW \x \bJ$
dynamo, which has recently been proposed \citep{rogachevskiietal03},
the $\delta^{(W)}$, $\kappa^{(W)}$, $\beta^{(D)}$ and
$\kappa^{(D)}$--effects are important. However, it turned
out that at least the simple model in plane geometry, which was used
to explain this proposal, fails to work as a dynamo as long as
$\delta^{(W)}$, $\kappa^{(W)}$, $\beta^{(D)}$ and $\kappa^{(D)}$ are
determined in the second--order correlation approximation
\citep{raeste05,ruedigeretal06}. In this context it is of interest
that the combination of the $\delta^{(W)}$, $\kappa^{(W)}$,
$\beta^{(D)}$ and $\kappa^{(D)}$--effects, as figure~\ref{WandD}b
shows, does not support the screw dynamo.
This is also in so far quite remarkable as we already know
that the $\beta^{(D)}$--effect supports it. It is the $\kappa^{(W)}$
and $\kappa^{(D)}$--effects which have a strong opposite influence.

Our results show that the total influence of the induction effects of the turbulence
on the threshold of the screw dynamo, which we have estimated above,
is weaker than the influences of some of the individual effects.
Obviously a part of them compensates each other.

\vspace{-1.0cm}

\vspace{1.0cm}

\section{Summary}
\label{summary}

The electromagnetic processes in Perm dynamo experiment are influenced
by turbulent motions accompanying the main flow.
Their influence on the dynamo process has been investigated within the framework
of mean--field electrodynamics.
The critical magnetic Reynolds number $R_U^*$ of the main flow,
which defines the onset of the screw dynamo,
grows with the magnetic Reynolds number $R_{u0}$ of the turbulence
and the also with the parameter $q$, that is, the ratio of the decay time of a magnetic field
extended over one correlation length to the correlation time.
Under the assumptions on these quantities made above, which we consider as realistic,
$R_U^*$ changes hardly by more than $0.5$ percent.
If they are by any reason larger, $R_U^*$ becomes larger, too.

The electromagnetic effects of the turbulence are analyzed in terms of $\alpha$,
$\gamma$, $\beta$, $\delta$ and $\kappa$--effects.
Despite helical structures in the turbulence there is no complete $\alpha$--effect.
The mean electromotive force due to this $\alpha$--effect has not always a component
in the direction of the mean magnetic field.
This fact explains the measurements made at an experimental setup
simulating the dynamo device in a smaller scale.
Despite this peculiarity the $\alpha$--effect supports the screw dynamo slightly.
The $\gamma$--effect acts always against the screw dynamo.
The two contributions to the $\beta$--effect, the $\beta^{(0)}$ and the $\beta^{(D)}$--effect,
act in opposite ways.
The $\beta^{(0)}$--effect leads to an mean-field diffusivity larger than the molecular one
and so raises the screw dynamo threshold.
The $\beta^{(D)}$--effect makes the mean--field diffusivity anisotropic
and strongly supports the screw dynamo.
Likewise the $\delta$--effects support the screw dynamo whereas the $\kappa$--effects act against it.
It is known that the $\delta^{(\Omega)}$ and $\kappa^{(\Omega)}$--effects in combination with
a proper, sufficiently strong rotational shear may work as a dynamo,
sometimes called $\bOmega \x \bJ$ dynamo.
They also support the screw dynamo.
Analogous to the $\bOmega \x \bJ$ dynamo the possibility of a $\bW \x \bJ$ dynamo has been discussed
in the literature,
for which the $\delta^{(W)}$, $\kappa^{(W)}$, $\beta^{(D)}$ and $\kappa^{(D)}$--effects are important,
but the question of its existence remained open so far.
Interestingly enough it was found that the combination of these effects does not support the screw dynamo.

\section*{Acknowledgment}

This work was partially supported by a grant of the BRHE Program and CRDF-009-0.
The authors thank Dr. Matthias Rheinhardt for inspiring discussions
and Dr. Franck Plunian for helpful comments.

\section*{Appendix}
\label{app}

The symmetric part $(\bnab \bV)^{(s)}$ of the gradient tensor of a vector field $\bV$,
in a Cartesian co--ordinate system defined by
$(\bnab \bV) ^{(s)}_{ij} = \frac{1}{2}(\p V_i / \p x_j + \p V_j / \p x_i)$,
has in a cylindrical co--ordinate system the components
\begin{eqnarray}
(\bnab \bV)^{(s)}_{rr} &=& \frac{\p V_r}{\p r} \, , \quad
    (\bnab \bV)^{(s)}_{r \varphi} = (\bnab \bV)^{(s)}_{\varphi r}
    = \frac{1}{2} \big(\frac{1}{r} \frac{\p V_r}{\p \varphi} +  \frac{\p V_\varphi}{\p r}
    - \frac{V_\varphi}{r} \big)
\nonumber\\
(\bnab \bV)^{(s)}_{r z} &=& (\bnab \bV)^{(s)}_{z r} = \frac{1}{2} \big(\frac{\p V_r}{\p z}
    + \frac{\p V_z}{\p r} \big) \, , \quad
    (\bnab \bV)^{(s)}_{\varphi \varphi} = \frac{1}{r} \big( \frac{\p V_\varphi}{\p \varphi} + V_r \big)
\label{eqa1}\\
(\bnab \bV)^{(s)}_{\varphi z} &=& (\bnab \bV)^{(s)}_{z \varphi}
    = \frac{1}{2} \big(\frac{\p V_\varphi}{\p z} + \frac{1}{r} \frac{\p V_z}{\p \varphi} \big) \, , \quad
    (\bnab \bV)^{(s)}_{zz} = \frac{\p V_z}{\p z} \, .
\nonumber
\end{eqnarray}


\bigskip

\begin{thebibliography}{10}

\bibitem[\protect\citeauthoryear{Avalos-Zuniga {\itshape{et al.}}}{2003}]{avalos-zunigaetal03}
R. Avalos-Zuniga, F. Plunian and A. Gailitis.
\newblock Influence of electromagnetic boundary conditions onto the onset of dynamo
action in laboratory experiments.
\newblock {\em Phys. Rev. E} 68, 066307, 2003.

\bibitem[\protect\citeauthoryear{Avalos-Zuniga and Plunian}{2005}]{avalos-zunigaetal05}
R. Avalos-Zuniga and F. Plunian.
\newblock Influence of inner and outer wall's electromagnetic properties on the onset of a stationary dynamo.
\newblock {\em Eur. Phys. J. B} 47, 127-135, 2005.

\bibitem[\protect\citeauthoryear{Denisov {\itshape{et al.}}}{1999}]{denisovetal99}
S.A.~Denisov, V.I.~Noskov, D.D.~Sokoloff, P.G.~Frik and
S.Yu.~Khripchenko.
\newblock On the possibility of laboratory realization of an unsteady MHD--dynamo.
\newblock {\em Doklady Mechanics} 44(4), 231--233, 1999.

\bibitem[\protect\citeauthoryear{Frick {\itshape{et al.}}}{2001}]{fricketal01}
P.~Frick, S.~Denisov, S.~Khripchenko, V.~Noskov, D.~Sokoloff and
R.~Stepanov.
\newblock A nonstationary dynamo experiment in a braked torus.
\newblock {\em Dynamo and Dynamics, a Mathematical Challenge},
edited by P.~Chossat, D.~Armbruster and I.~Oprea. Kluwer Dordrecht/Boston/London 2001,
pp.1--8.

\bibitem[\protect\citeauthoryear{Frick {\itshape{et al.}}}{2002}]{fricketal02}
P.~Frick, V.~Noskov, S.~Denisov, S.~Khripchenko, D.~Sokoloff,
R.~Stepanov and A.~Sukhanovski.
\newblock Non--stationary screw flow in a toroidal channel: way to a laboratory dynamo
experiment.
\newblock {\em Magnetohydrodynamics} 38, 136--155, 2002.

\bibitem[\protect\citeauthoryear{Gailitis {\itshape{et al.}}}{2000}]{gailitisetal00}
A.~Gailitis, O.~Lielausis, S.~Dement'ev, E.~Platacis, A.~Cifersons,
G.~Gerbeth,
  T.~Gundrum, F.~Stefani, M.~Christen, H.~H{\"a}nel and G.~Will.
\newblock Detection of a flow induced magnetic field eigenmode in the {R}iga
  dynamo facility.
\newblock {\em Phys. Rev. Lett.} 84, 4365--4368, 2000.

\bibitem[\protect\citeauthoryear{Gailitis {\itshape{et al.}}}{2001a}]{gailitisetal01}
A.~Gailitis, O.~Lielausis, E.~Platacis, S.~Dement'ev, A.~Cifersons,
G.~Gerbeth,
  T.~Gundrum and F.~Stefani.
\newblock Magnetic field saturation in the {R}iga dynamo experiment.
\newblock {\em Phys. Rev. Lett.} 86, 3024--3027, 2001.

\bibitem[\protect\citeauthoryear{Gailitis {\itshape{et al.}}}{2001b}]{gailitisetal01b}
A.~Gailitis, O.~Lielausis, E.~Platacis, G.~Gerbeth and F.~Stefani.
\newblock Riga dynamo experiment.
\newblock {\em Dynamo and Dynamics, a Mathematical Challenge},
edited by P.~Chossat, D.~Armbruster and I.~Oprea. Kluwer Dordrecht/Boston/London 2001,
pp.9--16.

\bibitem[\protect\citeauthoryear{Gailitis {\itshape{et al.}}}{2002a}]{gailitisetal02}
A.~Gailitis, O.~Lielausis, E.~Platacis, S.~Dement'ev, A.~Cifersons,
G.~Gerbeth, T.~Gundrum, F.~Stefani, M.~Christen and G.~Will.
\newblock Dynamo experiments at the {R}iga sodium facility.
\newblock {\em Magnetohydrodynamics} 38, 5--14, 2002.

\bibitem[\protect\citeauthoryear{Gailitis {\itshape{et al.}}}{2002b}]{gailitisetal02b}
A.~Gailitis, O.~Lielausis, E.~Platacis, G.~Gerbeth and F.~Stefani.
\newblock On the back--reaction effects in the {R}iga dynamo experiment.
\newblock {\em Magnetohydrodynamics} 38, 15--26, 2002.

\bibitem[\protect\citeauthoryear{Gailitis {\itshape{et al.}}}{2003}]{gailitisetal03}
A.~Gailitis, O.~Lielausis, E.~Platacis, G.~Gerbeth and F.~Stefani.
\newblock The Riga dynamo experiment.
\newblock {\em Surveys in Geophysics} 24, 247--267, 2003.

\bibitem[\protect\citeauthoryear{Gailitis {\itshape{et al.}}}{2004}]{gailitisetal04}
A.~Gailitis, O.~Lielausis, E.~Platacis, G.~Gerbeth and F.~Stefani.
\newblock Riga dynamo experiment and its Theoretical background.
\newblock {\em Physics of Plasmas} to appear, 2004.

\bibitem[\protect\citeauthoryear{Krause and R\"adler}{1971}]{krauseetal71}
F.~Krause and K.-H.~R\"adler.
\newblock Elektrodynamik der mittleren Felder in turbulenten leitenden Medien
und Dynamotheorie.
\newblock {\em In R. Rompe and M. Steebeck (eds.),
Ergebnisse der Plasmaphysik und Gaselektronik},
Akademie--Verlag Berlin, 1971 pp. 1-154.

\bibitem[\protect\citeauthoryear{Krause and R\"adler}{1980}]{krauseetal80}
F.~Krause and K.-H.~R{\"a}dler.
\newblock {\em Mean--Field Magnetohydrodynamics and Dynamo Theory}.
\newblock Akademie--Verlag Berlin and Pergamon Press Oxford, 1980.

\bibitem[\protect\citeauthoryear{Moffatt}{1978}]{moffatt78}
H.~K.~Moffatt.
\newblock Magnetic Field Generation in Electrically Conducting Fluids.
\newblock Cambridge University Press Cambridge, 1978.

\bibitem[\protect\citeauthoryear{Moffatt and Proctor}{1982}]{moffattetal82}
H.~K.~Moffatt and M.R.E.~Proctor.
\newblock The role of the helicity spectrum function in turbulent dynamo theory.
\newblock {\em Geophys. Astrophys. Fluid Dyn.} 21, 265--283, 1982.

\bibitem[\protect\citeauthoryear{Noskov {\itshape{et al.}}}{2004}]{noskovetal04}
V.~Noskov,S.~Denisov, P.~Frick, D.~Khripchenko, D.~Sokoloff and
R.~Stepanov.
\newblock Magnetic Field Rotation in the Screw Gallium Flow.
\newblock {\em European Phys. J.} B 41, 561--568, 2004.

\bibitem[\protect\citeauthoryear{Ponomarenko}{1973}]{ponomarenko73}
Yu.~B.~Ponomarenko.
\newblock On the theory of the hydrodynamic dynamo.
\newblock {\em PMTF} 1973(6), 47--51, 1973.
\newblock In Russian.

\bibitem[\protect\citeauthoryear{R\"adler}{1969}]{raedler69}
K.-H.~R\"adler.
\newblock \"Uber eine neue M\"oglichkeit eines Dynamomechanismus in turbulenten
leitenden Medien.
\newblock {\em Monatsber. Dtsch. Akad. Wiss. Berlin} 11, 272--279, 1969.

\bibitem[\protect\citeauthoryear{R\"adler}{1980}]{raedler80}
K.-H.~R\"{a}dler.
\newblock Mean-field approach to spherical dynamo models.
\newblock {\em Astron. Nachr.} 301, 101-129, 1980.

\bibitem[\protect\citeauthoryear{R\"adler}{1986}]{raedler86}
K.-H.~R\"adler.
\newblock Investigations on spherical kinematic mean--field dynamo models.
\newblock {\em Astron. Nachr.} 307, 89--113, 1986.

\bibitem[\protect\citeauthoryear{R\"adler}{2000}]{raedler00}
K.-H.~R\"adler.
\newblock The generation of cosmic magnetic fields
\newblock {\em In:} D. Page and J. Hirsch (eds.),
From the Sun to the Great Attractor,
1999 Guanajuato Lectures on Astrophysics,
\newblock {\em Lecture Notes in Physics},
Springer 2000, pp. 101-172.

\bibitem[\protect\citeauthoryear{R\"adler {\itshape{et al.}}}{2003}]{raedleretal03}
K.-H.~R\"adler, N.~Kleeorin and I.~Rogachevskii.
\newblock The mean electromotive force for MHD turbulence:
The case of weak mean magnetic field and slow motion.
\newblock {\em Geophys. Astrophys. Fluid Dyn.} 97, 249--274, 2003.

\bibitem[\protect\citeauthoryear{R\"adler and Stepanov}{2005}]{raeste05}
K.-H.~R\"adler and R.~Stepanov.
\newblock On the mean electromotive force due to turbulence
of a conducting fluid in the presence of a mean motion.
\newblock {\em Phys. Rev. E} 2006, in print. {\em arXiv:physics/0512120 }

\bibitem[\protect\citeauthoryear{Roberts}{1972}]{roberts72}
P.~H.~Roberts.
\newblock Kinematic dynamo models.
\newblock {\em Phil. Trans. R. Soc. London} A 272, 663--703, 1972.

\bibitem[\protect\citeauthoryear{Rogachevskii and Kleeorin}{2003}]{rogachevskiietal03}
I.~Rogachevskii and N.~Kleeorin.
\newblock Electromotive force and large--scale magnetic dynamo in a turbulent flow
with a mean shear.
\newblock {\em Phys. Rev. E} 68, 036301/1-12, 2003.

\bibitem[\protect\citeauthoryear{R\"udiger and Kitchatinov}{2006}]{ruedigeretal06}
G.~R\"udiger and L.~L.~Kitchatinov.
\newblock Vorticity and magnetic field generation in turbulent plane shear flows.
\newblock {\em Submitted to Astron. Nachr.}, 2006.

\bibitem[\protect\citeauthoryear{Schlichting}{1964}]{schlichting64}
H.~Schlichting.
\newblock Grenzschicht--Theorie.
\newblock G. Braun, Karlsruhe 1964.

\bibitem[\protect\citeauthoryear{Stepanov}{2000}]{stepanov00}
R.~Stepanov.
\newblock Study of the structure and the generation mechanism of galactic magnetic fields.
\newblock PhD thesis, ICMM, Perm, 2000. In Russian.


\end{thebibliography}
\end{document}